\documentclass[
reprint
,aps
,amssymb
,amsmath
,superscriptaddress
]{revtex4-1}

\usepackage{graphicx}
\usepackage{subcaption}
\usepackage{dcolumn}
\usepackage{fancyref}
\usepackage{bm}
\usepackage{xcolor}
\usepackage[toc,page]{appendix}
\usepackage[colorlinks=true,allcolors=blue]{hyperref}
\usepackage[normalem]{ulem}
\usepackage[fleqn]{mathtools}

\newcommand{\mdc}[1]{{\color{blue}#1}}
\newcommand{\amw}[1]{{\color{orange}#1}}

\newcommand{\comment}[1]{{\color{red}#1}}

\newcommand{\vvec}[1]{\mathbf{#1}}
\newcommand{\ttens}[1]{\mathrm{#1}}

\newcommand{\init}[1]{#1_{0}}

\def\sheetCoordOne{X_1}
\def\sheetCoordTwo{X_2}
\def\sheetCoords{X_1, X_2}
\def\sheetPartialOne{\partial_1}
\def\sheetPartialTwo{\partial_2}

\def\posr{\vvec{R}} 

\def\foldAngle{\theta}

\def\lengthscale{\zeta}

\def\mechScalar{\theta} 
\def\strain{\varepsilon} 
\def\mechStrain{\strain_{\mechScalar}}

\def\unsymStrain{S}
\def\metric{g}

\newcommand{\latticeVec}[1]{\vvec{l}^{(#1)}}
\newcommand{\latticeVecHat}[1]{\hat{l}^{(#1)}}
\def\normal{\vvec{N}}

\def\curvature{h}
\def\twist{t}
\def\bend{b}
\def\thirdCurvature{\kappa_3}
\def\normal{\hat{\vvec{N}}}
\def\scalarCurvature{R}
\def\riemann{R}
\def\christo{\Gamma}
\def\gaussCurvature{\kappa}

\def\coarseRot{\phi}

\def\principal{\lambda}   
\def\principalOne{\principal_1}
\def\principalTwo{\principal_2}

\def\poissonIn{\nu_{\text{in}}}   
\def\poissonOut{\nu_{\text{out}}}
\def\poissonMater{\nu_{\text{sheet}}}
\def\gamgam{\gamma} 
\def\aA{A}

\def\vertIndex{\mu}

\def\panelIndex{p}
\def\creaseIndex{q}
\def\bondIndex{uv}
\def\bondLength{d}
\def\bondIndexNodeOne{u}
\def\bondIndexNodeTwo{v}

\def\dihedral{\rho}

\def\panelDimOneTwo{a}
\def\panelDimOne{a_1}
\def\panelDimTwo{a_2}
\def\panelDimThree{c}

\def\levi{\epsilon} 

\def\ww{w} 

\def\curvatureVar{H}
\def\strainVar{\varepsilon}

\def\curvatureMode{p}
\def\curvatureModeOne{\curvatureMode_1} 
\def\curvatureModeTwo{\curvatureMode_2} 

\def\strainMode{q}
\def\strainModeOne{\strainMode_1} 
\def\strainModeTwo{\strainMode_2} 

\newcommand{\target}[1]{\tilde{#1}}
\newcommand{\cc}[1]{#1^*}
\newcommand{\data}[1]{{#1}^{(\text{obs})}}
\newcommand{\projection}[1]{{#1}^{(\text{proj})}}

\def\unitConversionFactor{\zeta}

\def\mod{k}
\def\modStretch{\mod_s}
\def\modBend{\mod_p}
\def\modFold{\mod_f}
\def\sheetThickness{l_t}
\def\ellRatio{L^*}
\def\barArea{a_{\text{bar}}}

\def\energyStretch{E_s}
\def\energyBend{E_p}
\def\energyFold{E_f}

\def\bendingModChris{K_b}

\def\mpPolar{\psi}
\def\mpAzimuth{\phi}

\begin{document}

\title{Orisometry formalism reveals duality and exotic nonuniform response in origami sheets}

\begin{abstract}

Origami metamaterial design enables drastic qualitative changes in the response properties of a thin sheet via the addition of a repeating pattern of folds based around a rigid folding motion. 
Known also as a mechanism, this folding motion will have a very small energy cost when applied uniformly; and yet uniform activation of such remains highly difficult to observe, these sheets instead generically displaying nonuniform response patterns which are not yet well understood. 
Here, we present a purely geometric continuum theory which captures the nonuniform, nonlinear response to generic loading as composed locally of the planar mechanism, as well as previously identified ``twist'' and ``bend'' modes which enable the patterned sheet to curve out of the plane across long distances.
Our numerical analysis confirms that these three modes govern the observed nonuniform response, varying smoothly across the sheet according to three PDEs which guarantee compatibility. In analogy with the recently solved case of planar mechanism metamaterials, these ``Orisometries'' (origami + isometries, so named by us) are subextensive but infinite in number, with each mode displaying, in the linear limit, ``sheared analytic'' spatial patterns which are controlled by the Poisson's ratio of the uniform folding mechanism. Furthermore, the ``planar'' mechanism-based deformation patterns superimpose with a mathematically dual space of ``non-planar'' twist/bend deformations to span the available soft linear response. Together, our findings furnish the first quantification of the number of soft response modes available, as well as the first intuitive quantification of their spatial distribution.


\end{abstract}

\author{Michael Czajkowski}
\affiliation{Dept. of Physics, Georgia Institute of Technology}
\author{James McInerney}
\affiliation{Dept. of Physics, University of Michigan, Ann Arbor, MI 48109, USA}
\author{Andrew M. Wu}
\affiliation{Dept. of Physics, Georgia Institute of Technology}
\author{D. Zeb Rocklin}
\affiliation{Dept. of Physics, Georgia Institute of Technology}
\date{\today}

\maketitle

\section{Introduction}

A conventional uncreased sheet of paper may be readily deformed into cylindrical and conical shapes with minimal applied force, but is not so easily stretched or sheared along its plane. This familiar behavior does not arise due to the specific composition of the paper, but rather is a geometric property which extends beyond paper: the thinner an elastic sheet, the more difficult it quickly becomes to stretch in comparison to bending~\cite{Jellett1849}. 
Deformations composed of bending without (in-plane) strain dominate the response properties of thin sheets; these are known as \emph{isometries}, the study of which extends over a century and a half~\cite{Jellett1849, Cerda1998, Witten2007} illuminating the otherwise cryptic nonlinear deformations that are observed.

\begin{figure}[!t]  
\begin{center}
    \includegraphics[width=0.48\textwidth]{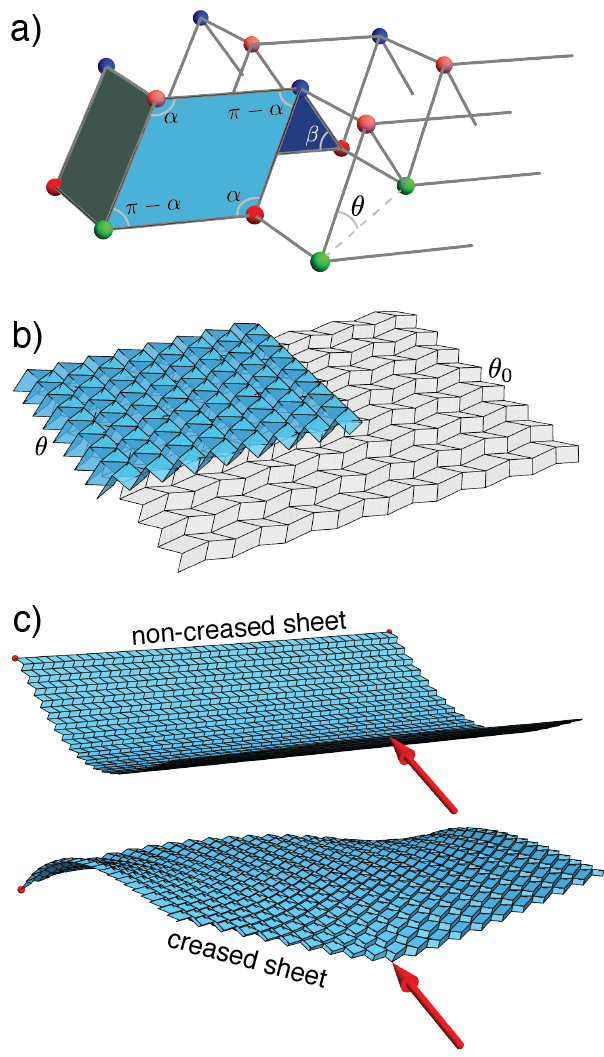}
    \caption{ \textbf{(a)} The ``Morph'' family of crease patterns are defined by two parallelograms, at angles $\alpha$ and $\beta$, and admit a quasi-planar folding motion \textbf{(b)}, parameterized by the angle $\foldAngle$ which, compared to the initial reference folding state at $\init{\foldAngle}$, enables apparent stretching of the surface the sheet approximates. \textbf{(c)} Two sheets under identical displacement loading (red arrows) and with the opposite corners held in place (red dots). In comparison to an uncreased sheet (top), a sheet which is creased according to a folding mechanism, (bottom) will display drastically different nonlinear nonuniform response, whose coarse shape change does not resemble any smooth isometry.
    \label{fig:main1}}
   \end{center}
\end{figure}

More recently, precise patterns of folding (i.e. Fig.~\ref{fig:main1}a) have been revealed which endow a conventional sheet with unconventional stretching capabilities~\cite{MIURA1985, Tachi2009, Schenk2013, Pratapa2019, Dieleman2019, Feng2020, mcinerney2022discrete}. This is achieved via the design based on a rigid folding \emph{mechanism}: an elaborate pathway of deformation which costs zero energy in the limit that the creases act as perfect frictionless hinges. When one ignores the fine details of the folding pattern in favor of the coarse shape changes of the sheet, such carefully corrugated sheets appear to subvert the inextensibility condition of isometric deformations: the sheets readily extend and change shape in the plane that they approximate (see Fig.~\ref{fig:main1}b). It is important to note that, while this mechanism folding motion is in fact isometric in the details of the crease microstructure, this is not a smoothly differentiable isometry and the motion leads to highly non-isometric behavior in the coarse macroscopic picture. 
As a result, the static response of a mechanism-based creased sheet is in sharp qualitative contrast with the response of the uncreased counterpart under identical loading (see Fig.~\ref{fig:main1}c). 
It then seems, in this coarse description, that the mathematics of the smooth isometries will no longer form a useful description of the lowest energy response available to a sheet with such a crease pattern.
What then is the spatial nature of these origami-based soft deformations? Do they admit a differentiable continuum mechanics description?





Answers to these questions reach considerably beyond individual folded sheets. Similar to the case with isometries, the exotic response properties of patterned origami sheets tend to arise from geometry rather than from the specifics of scale or material composition~\cite{ Mahadevan2005, Audoly2008, Schenk2013, Christensen2015, Bertoldi2017}. Such geometric insights into these materials are therefore particularly powerful because they will apply to origami at the submicroscopic scale of molecule just as well as the scale of a skyscraper~\cite{Brittain2001, Temmerman2007, Randhawa2011, Rogers2016, Li2019}. As a result, these geometric approaches have enabled a broad variety of applications including in robotics\cite{Okuzaki2008, Hawkes2010, Randhawa2011, Jeong2018, Lazarus2019, Son2022}, surgery~\cite{An2012, Felton2014, Miyashita2016}, and even in the development of deployable shelters for the unhoused~\cite{Temmerman2007}.

Despite all the advances they have enabled, the mechanisms themselves do not tell the complete story of geometry's role in origami response. For instance, although the quasi-planar mechanism motion programmed into these origami sheets is ideally zero energy, it remains quite difficult to uniformly activate~\cite{Misseroni2022}. Instead, a variety of nonuniform responses are observed~\cite{Misseroni2022}, and unreliable folding is pervasive even beyond these mechanisms~\cite{Waitukaitis2015, Stern2017, Chen2018, Misseroni2022, Trimble2022}.

Meanwhile, a recent series of investigations has begun to establish that for any uniform planar mechanism there must exist an associated space of nonuniform deformation modes which also cost zero energy in the continuum limit (i.e. with infinitesimally fine folding structure)~\cite{Czajkowski2021-2, Czajkowski2022, Czajkowski2022-2, Zheng2022}. This suggests that paired with the quasi-planar mechanisms, such as those of the the canonical Miura-ori and eggbox patterns, there will atleast be an affiliated space of quasi-planar nonuniform soft modes. In addition to these modes, it is well established that ubiquitous ``twist'' and ``bend'' modes enable such mechanism-based folded sheets to curl over long distances into nontrivial three dimensional shapes, again at zero energy cost in the continuum limit. These additional geometric effects have been utilized by Nassar et al.~\cite{Nassar2017, Nassar2022} to identify a striking variety of nontrivial, nonuniform soft deformations, and Xu et al.~\cite{xu2023derivation} have contributed further analytic configurations along with proof of softness and a continuum mechanics theory. However, a complete theory revealing the full spectrum of soft response in mechanism-based origami sheets is lacking. And so we refine the question addressed in this work: Does there exist a mathematical formulation analogous to the isometries which characterizes the coarse response of these sheets?

Here, we answer this question by assembling these varied perspectives on mechanism-based origami sheets to reveal a unified theory controlling the available soft deformations therein. Just as isometries of uncreased sheets take control as the thickness becomes small, the geometric limit of finely-detailed folding patterns (i.e. the continuum limit) reveals the role of our \emph{orisometries} in the response of mechanism-based origami sheets. In this theory, the three previously identified low-energy modes available to the sheet control deformation at the local level of just a few unit cells, rather than uniformly across the entire sheet. This confines the sheet to a ``hyperribbon'' (so named by our use here) in the space of the sheet fundamental forms. Following related previous work~\cite{Nassar2017, Nassar2022}, we allow these modes to vary slowly across the sheet, subject to spatial constraints, which reveals an infinite but subextensive space of available soft response. Furthermore, we find that the small linear deformations in this theory divide into the (quasi-) planar motions and the non-planar motions which are not only independent, but are mathematically dual to one another. These deformations are then controlled by an exceptional point in the system Poisson's ratio which divides bulk versus boundary soft mode patterns according to the auxetic versus anauxetic folding motion character. Our analytic results are confirmed using a variety of numerical force-balancing simulations within a particularly useful yet still broad class of ``Morph''\cite{Pratapa2019} fold patterns, altogether furnishing a governing framework which has remarkable similarities to the recent investigations in planar unimode metamaterials~\cite{Czajkowski2022-2, Zheng2022}.


\section{Local soft motions of mechanism-based origami sheets}

\begin{figure}[!t]  
\begin{center}
    \includegraphics[width=0.48\textwidth]{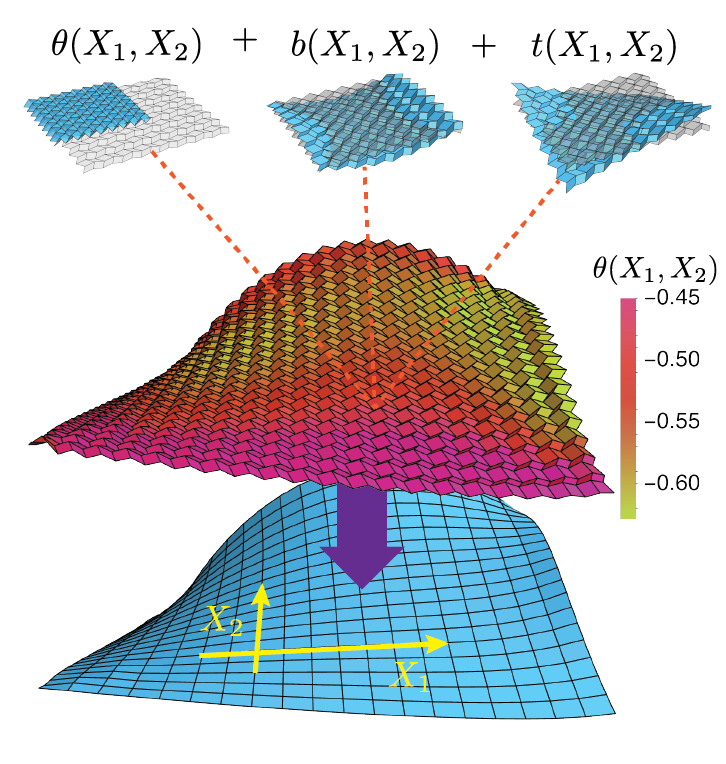}
    \caption{ The origami fold patterns locally have access to a planar mode (the mechanism), as well as a ``bend'' and a ``twist'' mode. Nonuniform patterns of these three modes comprise the nonuniform response to generic nonuniform loading, as approximated by the sheet midplane.
    }
    \label{fig:main2}
   \end{center}
\end{figure}

The investigation presented here concerns the ``Morph'' class of folding patterns~\cite{Pratapa2019}, illustrated diagrammatically in Fig.~\ref{fig:main1}a, and discussed in Appendix section~\ref{app:foldPatterns}  which includes the canonically studied examples of the Miura-ori and eggbox folds. We first enumerate the essential local ingredients of soft deformation which inform the subsequent analysis.
These folding patterns fall into the category of lattice origami, in which the undeformed folding pattern repeats itself upon translation by any integer combination of two lattice vectors $\init{\latticeVec{1}}, \init{\latticeVec{2}}$. For loading applied at lengthscales much longer than the unit cell size (the focus of this work), a coarse continuum description becomes useful. This is achieved by the origami sheet ``midplane'', depicted in Fig.~\ref{fig:main2} (bottom), which smoothly connects the unit cell centers. In this picture, the lattice vectors locally approximate the tangent vectors of the sheet, and travel along these vectors by continuous amounts $\sheetCoords$ allows the midplane surface $\posr(\sheetCoords)$ of the undeformed and corresponding deformed sheet to be parameterized. Specifically, this parameterization is defined so that $\latticeVec{i} = \partial_i \posr$ and differential geometric quantities such as the metric $\metric_{ij} \equiv \partial_i \posr \cdot \partial_j \posr$ and curvature $\curvature_{ij} \equiv -\partial_i \normal \cdot \partial_j \posr$ (arising from gradients in the surface normal $\normal(\sheetCoords)$) induced by the deformation may be estimated both numerically and analytically in the continuum limit approximation.


As the Morph patterns are contained within the recently studied ``four-parallelogram'' class~\cite{mcinerney2022discrete}, they each exhibit a nonlinear quasi-planar folding mechanism as displayed in Fig.~\ref{fig:main1}a. Quasi-planar means that this motion does not generate any curvature of the midplane surface. Instead, the mechanism motion induces a change in the surface metric away from the identity  
\begin{equation}\label{eq:metric}
\metric_{ij} = \frac{\latticeVec{i}(\foldAngle)\cdot \latticeVec{j}(\foldAngle)}{|\latticeVec{i}(\init{\foldAngle})| |\latticeVec{j}(\init{\foldAngle})|} \rightarrow \begin{bmatrix} \principalOne^2(\foldAngle) & 0 \\ 0 & \principalTwo^2(\foldAngle) 
\end{bmatrix}
\end{equation}
according to the change in the folding parameter $\foldAngle$ away from an initial reference (rest) value $\init{\foldAngle}$ shown in Fig.~\ref{fig:main1}a,b. Note that the surface metric and curvature tensors capture what are known as the first and second fundamental forms, and we use the language interchangeably in this work. Here, we have introduced the principal stretches $\principalOne, \principalTwo$, known from finite strain theory, which here capture the relative extension of the lattice vectors so that $\latticeVec{1}(\foldAngle) = \principalOne(\foldAngle)*\latticeVec{1}(\init{\foldAngle})$; this description accommodates the nonlinear changes such folded sheets are capable of. Note that the expression in this form is specific to the symmetries of the Morph patterns, and the metric need not remain diagonal along the mechanism motion in all cases. Furthermore, considering a small increment of the mechanism folding allows us to identify the planar Poisson's ratio
\begin{equation}\label{eq:poissonIn}
    \poissonIn(\init{\foldAngle}) \equiv - \frac{\principalTwo'(\init{\foldAngle})}{\principalOne'(\init{\foldAngle}) }
\end{equation}
at the chosen reference state of this mechanism.

\begin{figure}[!b]  
\begin{center}
    \includegraphics[width=0.48\textwidth]{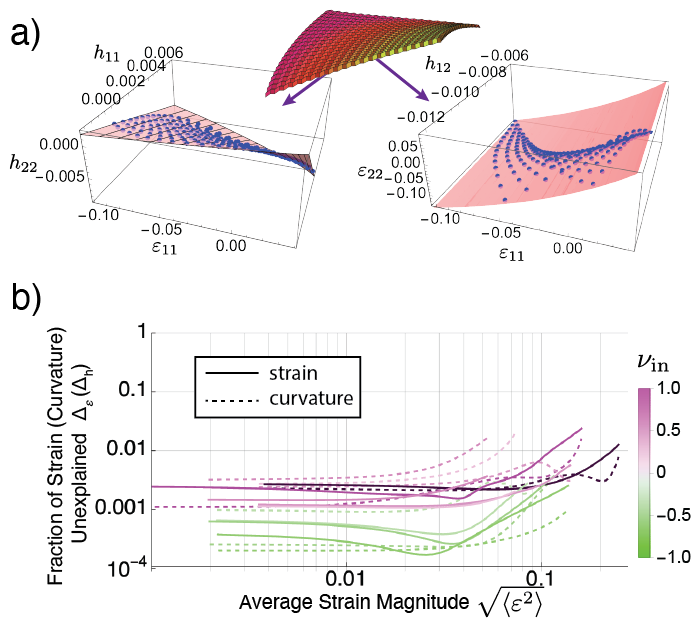}
    \caption{ \textbf{(a)} In the space of fundamental forms that describe the sheet deformation, response is confined to the ``hyperribbon'' we introduce here, with one long direction along the nonlinear planar mode and two short (linear) directions along the bend and twist modes. This is illustrated in two slices of the six dimensional fundamental form space for the numerically deformed configuration above. \textbf{(d)} Quantified root-mean-square distance from the hyperribbon (normalized by the root-mean-square spread \textit{along} the hyperribbon) in both the strain (solid lines) and curvature (dashed lines) subspaces across a variety of fold patterns with different rest state Poisson's ratios $\poissonIn$. Due to the normalization, the smallness of these quantities on an absolute scale indicates a high quality of adherence to the hyperribbon, which continues as the average strain is amplified to nonlinear values. \label{fig:main3} }
   \end{center}
\end{figure}

When the origami sheet panels are perfectly rigid, The quasi-planar motion is the only possible deformation. However, for typical origami sheet material which is thin and flexible, additional isometric deformations allow each panel to bend along the diagonals. This enables two additional modes of deformation, referred to as the ``twist'' and ``bend'' modes, which are known to only influence the second fundamental form
of the sheet~\cite{mcinerney2022discrete}. These modes, described in detail in Appendix.~\ref{app:coarseModes} only extend to linear order before generating energy-costly stretching of panels and may be superimposed together to generate the coarse curvature of the sheet 
\begin{equation}  \label{eq:curvature}
    \curvature_{ij}  \rightarrow \bend 
    \begin{bmatrix}
        \principalOne(\foldAngle) \principalOne'(\foldAngle) & 0 \\
        0 & -\principalTwo(\foldAngle) \principalTwo'(\foldAngle)
    \end{bmatrix}
    + \twist
    \begin{bmatrix}
        0 & 1 \\ 1 & 0
    \end{bmatrix} \, ,
\end{equation}
where $\bend$ and $\twist$ are amplitudes of the bend and twist modes respectively. The expression for the bending curvature in Eq.~\ref{eq:curvature} is determined from the established result that the bending mode generates an out-of-plane Poisson's ratio $\poissonOut \equiv -\frac{\curvature_{22}}{\curvature_{11}}$ which is equal and opposite to the in-plane ratio $\poissonOut = -\poissonIn$~\cite{Schenk2013, Wei2013, Pratapa2019, mcinerney2022discrete}.


\section{Origami sheet deformations adhere to a shape ``hyperribbon''}

When applied uniformly, each of the three modes established in the previous section incur only a small energy cost proportional to the crease folding stiffness $\modFold$ and panel bending stiffness $\modBend$ without panel stretching $\modStretch$. As discussed in greater detail in Appendix Section.~\ref{app:realParams}, the geometry of the thin sheet dictates a qualitative difference between these moduli $\modStretch \gg \modFold,  \modBend$, and hence a qualitative difference between these soft modes and other arbitrary uniform deformations which include panel stretching. However, promoting these soft mode amplitudes $\foldAngle, \twist, \bend$ to scalar fields $\foldAngle(\sheetCoords), \twist(\sheetCoords), \bend(\sheetCoords)$  and allowing these modes to vary slowly across the sheet may impose an additional energy cost (see Fig.~\ref{fig:main2}). For very slow gradients, the additional cost must be proportional to the ratio of the unit cell size $\sim a$ to the lengthscale of variation of the mode $\lengthscale$. It follows that in the continuum limit where $\lengthscale \gg a$, this additional energy from spatial variations must go to zero and such nonuniform soft mode deformations will again have (low) energy proportional to the soft crease and panel stiffnesses. Whenever possible, the sheet will therefore use these slowly varying soft modes to respond to loading, rather than more energy-costly deformations which stretch the constitutive sheet.


According to the rigidity theorem of Bonnet~\cite{toponogov2006differential}, the shape of any smooth surface is completely described by the first and second fundamental forms (captured by the components of the metric and curvature tensor, respectively) across the sheet. Each point on the surface may be assigned a 6-dimensional coordinate (3 coordinates for each of the two symmetric tensors), and the entire sheet may be interpreted as a volume in this 6-dimensional space of these forms.  Therefore, a smooth differentiable surface may be viewed as a continuous volume extending in the 6-dimensional space of the components of these forms, each point on the surface having coordinates ($\metric_{11}, \metric_{12}, \metric_{22}, \curvature_{11}, \curvature_{12}, \curvature_{22}$). However, for the response of an origami sheet, the lowest energy possibilities will be described by the amplitudes of planar($\foldAngle$), twist($\twist$), and bend($\bend$) modes. Therefore, such response must be confined to a 3 dimensional manifold within the 6-dimensional fundamental form space. As this surface is parameterized by one nonlinear number ($\foldAngle$) and two small linear numbers ($\twist, \bend$), we refer to such as the \textit{hyperribbon}. 

\begin{figure}[!t]  
\begin{center}
    \includegraphics[width=0.48\textwidth]{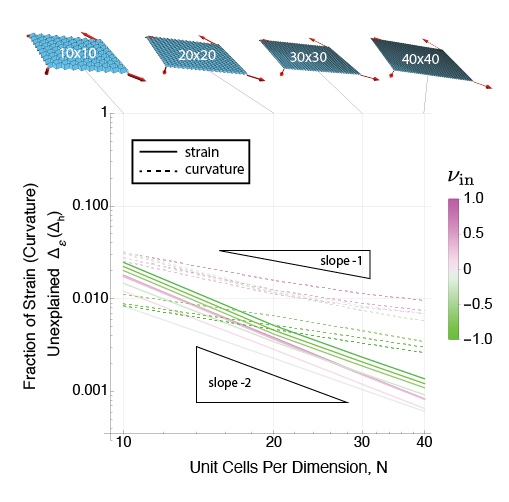}
    \caption{ The averaged fractional deviation from the hyperribbon is very small and ubiquitously improves (becomes smaller) as the system size is increased and the continuum limit is approached. The error in the curvature (dashed lines) scales approximately with the lattice coarseness ($1/N$), while the error in the metric (solid lines) scales approximately with the square coarseness ($1/N^2$).  \label{fig:main4} }
   \end{center}
\end{figure}

To evaluate the adherence of real origami sheets to their shape hyperribbon, as well as the subsequent analyses herein, we numerically investigate force-balanced deformations using the MERLIN2 simulation framework for an ``enhanced'' bar-hinge model of the origami sheet~\cite{Filipov2017, Liu2017, liu2018highly}. To facilitate algorithmic convergence, the data displayed in Figs.~\ref{fig:main3},\ref{fig:main4}\&\ref{fig:main5} are obtained under arbitrary point-loading conditions and using ``rigid'' parameters in which the crease folding incurs no energy penalty ($\modFold \rightarrow 0$). In contrast, to better represent real world response the data illustrated in Figs.~\ref{fig:main1}\&\ref{fig:main2} are obtained using ``realistic'' parameters emulating a constitutive sheet having thickness $1\%$ of the panel dimension (i.e. roughly capturing printer paper), and thereby having a small, but finite crease folding modulus $\modFold$. All of these numerical results, as well as the applied loading conditions, and simulation protocol are described in Appendix.~\ref{app:simulations}.

As shown in Fig.~\ref{fig:main3}, the  numerically force-balanced \emph{nonlinear} deformations of a variety of these origami sheets exhibit very close adherence to this hyperribbon ($<1\%$ error) out to strains of $10-20\%$. As suggested in the energetic argument above, the adherence to the hyperribbon ubiquitously improves as the continuum limit is approached in Fig.~\ref{fig:main4}. This non-hyperribbon error follows a suggestive power-law scaling, in which the curvature error scales inversely with the system size, while the metric error improves faster, with the scaling $\sim N^{-2}$. 


\section{Field theory for origami sheets from compatibility}

The nonuniform soft response of the origami sheet will be composed from these three spatially varying modes  $\foldAngle(\sheetCoords), \twist(\sheetCoords), \bend(\sheetCoords)$. However, it is well-known in the study of differential geometry that not all spatial patterns of the metric and curvature tensors correspond to any realizable (i.e. immersible) surface: regardless of deformation mechanics and energy, geometric compatibility restricts the spatial evolution of these tensors across the surface. Therefore, as the soft modes $\foldAngle(\sheetCoords), \twist(\sheetCoords), \bend(\sheetCoords)$ control the metric and curvature tensors in our origami sheet, they are in turn governed by geometric compatibility conditions. 
Again relying on the rigidity theorem of Bonnet, it is known that the Gauss-Codazzi-Mainardi-Peterson equations are both necessary and sufficient to guarantee such a geometrically valid sheet~\cite{toponogov2006differential}. These equations, discussed in detail in Appendix~\ref{app:gaussCodazzi} constitute a set of three PDEs, and with Eqs.~\ref{eq:metric}\&\ref{eq:curvature} the general form of these PDEs may be reduced to a form specific to this category of origami sheet. In full form, these equations are:
\begin{align}\nonumber
    \frac{1}{\principalOne \principalTwo} & \levi^{ij} \partial_i \left( B_{jk}(\foldAngle) \partial_k \foldAngle  \right) \\ \label{eq:gauss}
    & =  - \frac{1}{\principalOne^2 \principalTwo^2} \left[  \bend^2 \principalOne' \principalTwo' + \twist^2 \right] \\ \nonumber
    0 = & [\sheetPartialOne \twist - \principalOne' \sheetPartialTwo         \bend - \bend \principalOne'' \sheetPartialTwo \foldAngle] + \twist (\frac{\principalTwo'}{\principalTwo} - \frac{\principalOne'}{\principalOne}) \sheetPartialOne \foldAngle  \\ \label{eq:codazziOne}
     &  + b \left( \frac{\principalOne'}{\principalOne^2} -  \frac{\principalTwo'}{\principalTwo^2} \right) * \principalOne \principalOne' \sheetPartialTwo \foldAngle  \\ 
    \nonumber
    0 = & -[\sheetPartialTwo t + \principalTwo' \sheetPartialOne b + b \principalTwo'' \sheetPartialOne \foldAngle] + t (\frac{\principalTwo'}{\principalTwo} - \frac{\principalOne'}{\principalOne}) \sheetPartialTwo \foldAngle  \\ \label{eq:codazziTwo}
    &  - b \left( \frac{\principalOne'}{\principalOne^2} -  \frac{\principalTwo'}{\principalTwo^2} \right) * \principalTwo \principalTwo' \sheetPartialOne \foldAngle 
     \, ,
\end{align}
where
\begin{equation}\label{eq:bmatrix}
    \ttens{B} = 
    \begin{bmatrix}
         0 & -\frac{\principalOne'}{\principalTwo} \\
         \frac{\principalTwo'}{\principalOne} & 0
    \end{bmatrix} 
    \, .
\end{equation}
%
The first (Eq.~\ref{eq:gauss}) of these arises from the Gauss equation and is second order, while Eqs.~\ref{eq:codazziOne}\&\ref{eq:codazziTwo} are the first order Codazzi equations. While this set of nonlinear PDEs is highly nontrivial to solve, the balance of the number of fields and equations suggests that boundary conditions will be sufficient to determine the soft mode patterning through the bulk of an origami sheet. Despite the natural exceptions to this, it therefore appears that the number of soft nonuniform deformations available to the origami sheet globally will be infinite but subextensive, similar to the planar unimode case~\cite{Czajkowski2022, Zheng2022, Czajkowski2022-2}. Note that in the absence of the twist and bend fields, this reduces as expected to the planar unimode compatibility theory established in Refs.~\cite{Czajkowski2022, Zheng2022, Czajkowski2022-2}.


\subsection{Dual spaces of sheared analytic functions capture linear deformations in the origami sheet}

\begin{figure}[!t]  
   \begin{center}
    \includegraphics[width=0.48\textwidth]{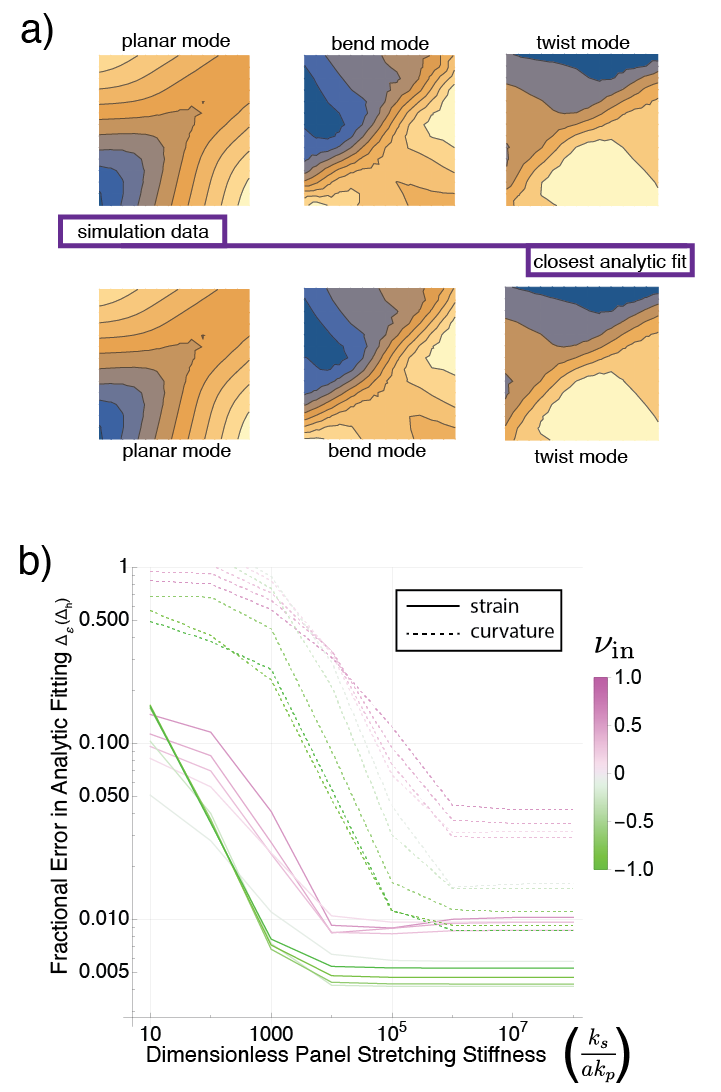}
    \caption{  Under linear deformations, the spatial distribution of soft mode amplitudes may be described using sheared analytic functions. \textbf{(a)} The nontrivial spatial patterns of the three modes are reproduced to high qualitative accuracy by closest fit sheared analytic function. \textbf{(b)} The accuracy of these fittings is a proxy for the degree of success of the soft mode theory. As expected, the error in fitting the metric (solid lines) and the curvature (dashed lines) becomes small and the theory becomes accurate only in the limit that the ratio of the panel stretching stiffness  to the panel bending stiffness becomes very large ($\sim 10^5$). \label{fig:main5} }
   \end{center}
\end{figure}

The theory in Eqs.~\ref{eq:gauss}-\ref{eq:codazziTwo} governs the nonuniform soft modes, yet analytic solutions are generally challenging to obtain. We therefore consider the limit of linearly small deviatoric mode amplitudes $\delta\foldAngle, \delta \twist, \delta \bend$,
where these equations may be written in the compact form
\begin{align}\label{eq:gauss_linear}
    0 & = \partial_\ww \partial_{\bar{\ww}} \delta \foldAngle \\ \label{eq:codazziOne_linear}
    0 & = \partial_{\bar{\ww}} \curvatureVar \\ \label{eq:codazziTwo_linear}
    0 & = \partial_{\ww} \bar{\curvatureVar}
\end{align}
where 
\begin{align}\label{eq:w}
    \ww & = \sheetCoordOne + \sqrt{\poissonIn}\sheetCoordTwo \\ \label{eq:wbar}
    \bar{\ww} & = \sheetCoordOne - \sqrt{\poissonIn}\sheetCoordTwo
\end{align}
are a transformed ``sheared'' set of spatial variables established in Ref.~\cite{Czajkowski2022-2} and
\begin{align}\label{eq:curvatureVar}
    \curvatureVar & = \twist - \sqrt{\frac{1}{\poissonIn}} \principalTwo'\bend \\ \label{eq:curvatureVarBar}
    \bar{\curvatureVar} & = \twist + \sqrt{\frac{1}{\poissonIn}} \principalTwo'\bend
\end{align}
describes the curvature. 
These new variables $\curvatureVar, \bar{\curvatureVar}$ determine the twist and bend fields which, in the hyperribbon description of the origami sheet, thereby determine the entire curvature tensor according to Eq.~\ref{eq:curvature}. Although energetic couplings between the metric and the curvature tensors are not ruled out, there is no coupling from geometric compatiblity in this linear limit, and the metric and curvature may be determined independently. Therefore, we may think of these deformations as composed of two steps: first a quasi planar soft deformation is applied, followed by a ``non-planar'' deformation, which does not contribute to the planar strains. 

Consider first these non-planar deformations. In the limit of pure dilational planar folding mode ($\poissonIn = -1$), the sheared coordinate transformation is the conventional complex-variable representation of the plane $\ww, \bar{\ww} \rightarrow z, z^*$, and according to Eqs.~\ref{eq:codazziOne_linear}\&\ref{eq:codazziTwo_linear}, the modes will be controlled by a conventional holomorphic $\partial_{\bar{z}}\curvatureVar =0$ (i.e. conformal, analytic) and anti-holomorphic $\partial_z\bar{\curvatureVar} = 0$ mapping. More generally (i.e. for arbitrary $\poissonIn$) the solutions for the soft curvatures may be written down using ``sheared analytic functions'' as introduced in Ref.~\cite{Czajkowski2022-2}. Similar to the dilational case, such functions are determined solely by one or the other of the sheared analytic variables $\ww, \bar{\ww}$ but not both, and in the present case we have $\curvatureVar(\ww, \bar{\ww}) \rightarrow \curvatureVar(\ww)$ and $ \bar{\curvatureVar}(\ww, \bar{\ww}) \rightarrow \bar{\curvatureVar}(\bar{\ww})$.  This formalism, which determines the spatial distribution of the soft modes, is described in greater detail in Appendix~\ref{app:linearModesDefine}. 
These spatial patterns are controlled by an exceptional point at $\poissonIn=0$, where the sheared spatial variables transition from being real-valued when $\poissonIn>0$ (anauxetic folding mechanism) to being complex-valued when $\poissonIn<0$ (auxetic folding mechanism). Tuning the system across this transition, the bulk oscillatory modes of the anauxetic systems transform into evanescent modes which decay away from the boundary in the auxetic systems.


The remaining question of planar folding mode compatibility may be again addressed directly relying on previous work. By virtue of decoupling from non-planar effects, the planar mode compatibility equation in Eq.~\ref{eq:gauss_linear} must be equivalent to the linear analysis of planar unimode metamaterials addressed in Ref.~\cite{Czajkowski2022-2}. 
As a result of this, the second Eq.~\ref{eq:gauss_linear} may be exactly rewritten as two first order PDE equations
\begin{align}\label{eq:gauss1}
    0 & = \partial_{\bar{\ww}} \strainVar \\ \label{eq:gauss2}
    0 & = \partial_\ww \bar{\strainVar}
\end{align}
where we will define $\coarseRot$ as the local coarse rotation of each unit cell in the sheet reference plane 
and $\strainVar  \equiv \delta\foldAngle {\principalOne'} - \frac{1}{\sqrt{\poissonIn}} \coarseRot$ and $\bar{\strainVar}  \equiv \delta\foldAngle {\principalOne'} + \frac{1}{\sqrt{\poissonIn}} \coarseRot$ are special defined variables capturing the planar components of deformation. 
Eqs.~\ref{eq:gauss1}\&\ref{eq:gauss2} are clearly identical equations to those controlling the non-planar modes (Eqs.~\ref{eq:codazziOne_linear}\&\ref{eq:codazziTwo_linear}), simply replacing the variables $\curvatureVar, \bar{\curvatureVar} \rightarrow \strainVar, \bar{\strainVar}$.
Therefore, the same sheared analytic functions used above to generate a pattern of the non-planar modes $\twist, \bend$, can be instead used to generate a pattern of the planar modes $\delta\foldAngle, \coarseRot$. Each non-planar deformation then corresponds precisely with one of the planar deformations, and vice versa, thereby establishing a mathematical duality, similar to that revealed in Ref.~\cite{Czajkowski2022-2}. 
Any arbitrary linear deformation of the origami sheet is now generated by choosing one set of sheared analytic functions controlling the planar modes and a second set of sheared analytic functions controlling the non-planar modes, and superimposing these deformations in the origami sheet.
It is important to note that in a given deformation the nonplanar mode need not be the corresponding dual pattern to the superimposed planar mode and vice versa: this represents a special case of a ``self-dual'' deformation, the physical implications of which will require further investigation. 

In Fig~\ref{fig:main5} the distribution of the fundamental forms along the sheet is described to high accuracy (fractional error only reaching as high as $5 \%$) using these sheared analytic modes when the sheet bending $\modBend, \modFold$ becomes much more energetically costly than sheet stretching $\modStretch$. This confirms the notion that mechanism-based origami composed of sufficiently thin sheets will have response governed by the orisometries presented herein. 



\section{Discussion}

In contrast to related previous investigations into nonuniform patterns of folding in lattice origami sheets~\cite{Nassar2017, Nassar2022, xu2023derivation}, the formalism presented here in terms of these spatially varying mode amplitudes, is unique. Our approach to the compatibility requirements which constrain these modes brings new qualitative insight to the breadth and spatial character of response.

This orisometries formalism reveals also that the soft deformations correspond in the linear limit to the sheared analytic modes introduced previously~\cite{Czajkowski2022-2}. These are governed by an exceptional point in the Poisson's ratio at $\poissonIn=0$, which qualitatively divides the character of the spatial response. Further, the in-plane and out-of-plane modes which superimpose to span the space of soft response are mathematically dual spaces, suggesting connections to other mechanical dualities~\cite{Pretko2018,Fruchart2020, Czajkowski2022-2}. While this gives great insight into the character of the soft modes available to these origami sheets, it also adds further emphasis to the question: what is the deeper principle that these dualities in mechanism-based metamaterials arise from?


While the analysis herein focuses on a particular mode-based formalism which serves our purposes, many long-established alternatives exist for the description of manifold deformation in three dimensions. Our analysis nonetheless indicates that any such formalism, in describing the low energy-deformations of such a sheet, must still exhibit this duality therein. This new duality compounds with the duality identified previously between the spatial patterns of soft strains and the stresses in planar unimode systems~\cite{Czajkowski2022-2}; it is expected that they will coexist in the origami sheet, yet an in-depth analysis of the constitutive laws that may control both the sheet strains and curvatures should be established first. The formalism of spatially varying soft mode variables introduced here will lend a useful framework for development of an energy functional to predict specific deformation response.

Interestingly, according to constraint counting, the presence of three local deformation modes is not unique to the patterns explored here. All fold patterns may be considered to be effectively triangulated with soft panel bending in this manner and, starting from a quasi-planar state, three modes must always be present and governed by the equations of compatibility~\cite{McInerney2020}. Therefore, while the Morph family of fold patterns explored here are chosen for analytic convenience, a subextensive but infinite space of soft modes is expected to be the norm rather than the exception for periodic fold patterns.
However, in such origami sheets which approximate a cone or a cylinder in their resting state, the mathematics of soft deformation must change. In particular, such screw-periodic geometries exhibit two linear isometries rather than the three for the spatially-periodic sheets considered herein. It will be interesting to explore the possible soft modes and the challenges that may arise in discretized descriptions of e.g. curvature in this setting.

The orisometry approach presented here also presents new challenges and questions. For instance, we pose the question of what governs the suggestive and distinct continuum limit scaling exponents in the hyperribbon fitting error shown in  Fig.~\ref{fig:main4}. Beyond this, our theory presents a series of nonlinear equations, and the solutions to these equations are largely unexplored and potentially rich with possibilities. Finally, in analogy with the phenomena of stress focusing and  developable cones in thin sheet response, these is a question of where and when this theory here may break down, requiring more singular and localized phenomena in place of smoothness. 



\onecolumngrid

\newpage

\appendix


\section{Morph-class of fold patterns}
\label{app:foldPatterns}

\begin{figure*}[!t]  
\begin{center}
    \includegraphics[width=0.97\textwidth]{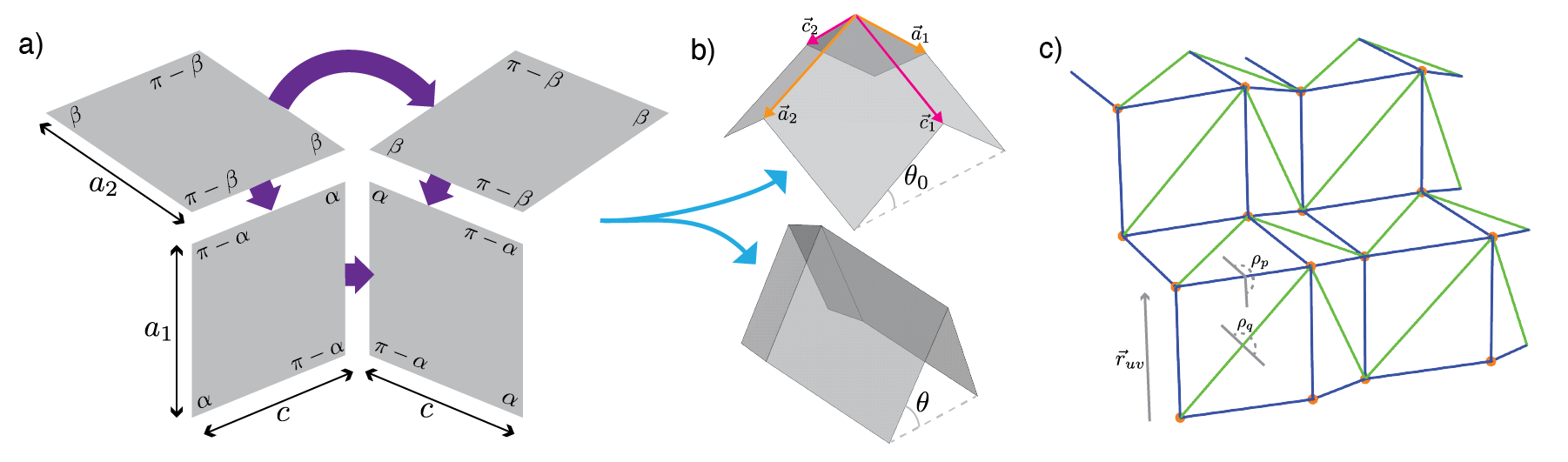}
    \caption{ \label{fig:app_morph} Construction and simulation of Morph patterns in a bar-hinge model. \textbf{(a)} The Morph unit cell is composed of four parallelogram panels, which come in indistinguishable pairs. Parameters to generate these panels, as well as the recipe to connect them into a unit cell illustrated here. \textbf{(b)} Once constructed, such a Morph pattern sheet is still able to change shape via the folding motion, parameterized by the folding $\foldAngle$ shown here for a single unit cell. \textbf{(c)} A representative model of such a sheet is achieved by adding diagonal crease lines across the parallelogram panels and adding energy penalties for changing the dihedral angles across these panel diagonals $\dihedral_\panelIndex$, changing the dihedral angles across the creases at the panel edges $\dihedral_\creaseIndex$ as well as spring energy penalties for stretching the crease lines along their axis $\vvec{r}_\bondIndex$ stretching from node $u$ to $v$. }
   \end{center}
\end{figure*}

In this work we focus on a class of fold patterns dubbed the ``Morph''~\cite{Pratapa2019}. Each pattern in this class is composed of a repeating pattern of four parallelogram panels (in two identical pairs) as shown in Fig.~\ref{fig:app_morph}. To generate an arbitrary member of this class of fold patterns, two ``sector angles'' $\alpha$ and $\beta$, and three edge lengths $\panelDimOne$, $\panelDimTwo$ and $\panelDimThree$ must be provided, as shown in Fig.~\ref{fig:app_morph}a. 
For rigid origami (i.e. panels can neither bend nor stretch) this information is only sufficient to determine the crease pattern of the origami sheet, but not to determine the stress free reference state. This is because the origami sheet has a single \emph{mechanism} degree of freedom, which extends nonlinearly, as shown in main text Fig.~\ref{fig:main1}. Such a mechanism, in the absence of any energy cost of folding along creases, provides a degenerate space of ground states which the sheet may realize.  

To understand this rigid mechanism folding motion, consider a single isolated vertex at the intersection of the four parallelogram panels (again, two pairs of identical parallelograms). We may consider the four dihedral angles of the creases between the panels to form a sufficient set of nontrivial degrees of freedom describing the vertex shape. These dihedral angles quantify folding, and for a particular folding configuration around a single vertex to be geometrically valid (i.e. to permit flat, unstretched panels between them), these angles must satisfy the well-known Belcastro-Hull compatibility conditions~\cite{Belcastro2002}. This compatibility requirement generically imposes three distinct nontrivial constraints on the dihedral angles connected to a single vertex. For the Morph folding patterns, four creases meet at the vertex and, accounting for the three constraints, one nonlinear degree of freedom is left over to vary the vertex configuration. This captures the rigid folding motion available to the sheet. We parameterize this folding motion using a single scalar parameter $\foldAngle$, as shown in Fig.~\ref{fig:app_morph}b. This vertex, along with the four panels it connects, defines a unit cell of the repeating fold pattern. It is, however, neither generic nor trivial that repeating copies of such a unit cell can be assembled together to form a single coherent origami sheet. The key restriction, as shown in Ref.~\cite{Waitukaitis2015, mcinerney2022discrete}, is that each panel must be chosen to be parallelogram shaped. This is, by definition, true of the family of Morph patterns considered here. With a bit of thought it is clear that the edges on opposite sides of the unit cell must match in both length and orientation, and the repeating crease pattern is composed of such a unit cell, translated by the lattice vectors and stitched together along crease lines. And, as this is possible for arbitrary choices of the folding angle $\foldAngle$ of the single unit cell, this rigid folding motion is preserved in the infinite sheet. 

Given the defining parameters $\panelDimOne, \panelDimTwo, \panelDimThree, \alpha, \beta$ and the rigid folding motion parameter $\foldAngle$, the procedure for generating a sheet configuration in 3d space is straightforward but analytically cumbersome~\cite{mcinerney2022discrete}. Conveniently, in the case of the Morph patterns, symmetry simplifies this analysis. For simplicity, we restrict our analysis to the case $\panelDimOne=\panelDimTwo \equiv \panelDimOneTwo$. Consider the four vectors (orange $\vvec{a}_1, \vvec{a}_2$ and pink $\vvec{c}_1, \vvec{c}_2$) emanating from a vertex, as shown in Fig.~\ref{fig:app_morph}b. These edge vectors may be used to assemble the lattice vectors $\latticeVec{1} = \vvec{c}_1 - \vvec{c}_2$ and $\latticeVec{2} = \vvec{a}_1 - \vvec{a}_2$ connecting identical vertices in neighboring unit cells. Because the Morph patterns are symmetric on mirroring through the $\latticeVec{1}$ direction, these lattice vectors are orthogonal for all choices of the folding angle. These vectors may then be used to establish a local orthonormal basis $\latticeVecHat{1}, \latticeVecHat{2}, \normal = \latticeVecHat{1} \times \latticeVecHat{2}$. The edge vectors may then be written in terms of their components along these basis elements via
\begin{align}\label{eq:app_edgeVectors}
    \vvec{a}_1 & = \panelDimOneTwo \left( \cos(\foldAngle) \latticeVecHat{2} - \sin(\foldAngle) \normal \right) \\    
    \vvec{a}_2 & = - \panelDimOneTwo \left( \cos(\foldAngle) \latticeVecHat{2} + \sin(\foldAngle) \normal \right) \\
    \vvec{c}_1 & = \panelDimThree \left(  \sin(\mpPolar) \cos(\mpAzimuth) \latticeVecHat{1} +  \sin(\mpPolar) \sin(\mpAzimuth) \hat{\latticeVec{2}} +  \cos(\mpPolar) \normal \right) \\
    \vvec{c}_2 & = \panelDimThree \left( - \sin(\mpPolar) \cos(\mpAzimuth) \latticeVecHat{1} +  \sin(\mpPolar) \sin(\mpAzimuth) \hat{\latticeVec{2}} +  \cos(\mpPolar) \normal \right)  \, ,
\end{align}
where we have implicitly defined a polar $\mpPolar$ and an azimuthal $\mpAzimuth$ angle to describe the $c$-vector orientations in spherical coordinates. Here, we have used the symmetry of the fold pattern to relate $\vvec{c}_1$ to $\vvec{c}_2$ as the mirror across the plane with normal $\latticeVecHat{1}$. The goal is to determine the changes to the edge orientations which will preserve the sector angles $\alpha, \beta$ between adjacent edges. Again this is simplified into two constraining conditions by the symmetry across the $\latticeVecHat{1}$ direction. We require
\begin{align} \nonumber
    \vvec{a}_1 \cdot \vvec{c}_1 & = \panelDimOneTwo \panelDimThree \cos(\alpha) \\ \nonumber
    \vvec{a}_2 \cdot \vvec{c}_1 & = \panelDimOneTwo \panelDimThree \cos(\beta)  \, ,
\end{align}
and, plugging in, these conditions becomes
\begin{align} \nonumber
    \panelDimOneTwo \panelDimThree \cos(\alpha) & =  \panelDimOneTwo \panelDimThree \left( \cos(\foldAngle) \sin(\mpPolar) \sin(\mpAzimuth) - \sin(\foldAngle) \cos(\mpPolar) \right) \\ \nonumber
    \panelDimOneTwo \panelDimThree \cos(\beta) & = -  \panelDimOneTwo \panelDimThree \left( \cos(\foldAngle) \sin(\mpPolar) \sin(\mpAzimuth) + \sin(\foldAngle) \cos(\mpPolar) \right) \, .
\end{align}
These are solved by the relations
\begin{align}\label{eq:app_angleRelation1}
    \cos(\mpPolar(\foldAngle, \alpha, \beta)) & =  -\frac{\cos(\alpha) + \cos(\beta)}{2 \sin(\foldAngle)} \\ \label{eq:app_angleRelation2}
    \sin(\mpAzimuth(\foldAngle, \alpha, \beta)) & =  -\frac{\cos(\alpha) - \cos(\beta)}{2 \cos(\foldAngle) \sin(\mpPolar(\foldAngle, \alpha, \beta))} \, .
\end{align}
Here, the geometric angle $\foldAngle$ is the chosen parameter of the rigid folding motion. Therefore, we view $\mpPolar(\foldAngle, \alpha, \beta)$ as a function of this angle and the sector angles that define the vertex, whose value is obtained by solving Eq.~\ref{eq:app_angleRelation1}. Similarly, $\mpAzimuth(\foldAngle, \alpha, \beta)$ may be viewed as a function determined by the vertex geometry and folding, with a value that can be determined by using $\mpPolar$ to solving Eq.~\ref{eq:app_angleRelation2}. 

Importantly, these geometric relations allow us to write the lattice vectors analytically in terms of the fold angle and sector angles
\begin{align}\label{eq:app_latticeVecMorphTwo}
    \latticeVec{2} & = 2  \panelDimOneTwo \cos(\foldAngle) \latticeVecHat{2} \\ \label{eq:app_latticeVecMorphOne}
    \latticeVec{1} & = 2  \panelDimThree \sin(\mpPolar(\foldAngle, \alpha, \beta)) \cos(\mpAzimuth(\foldAngle, \alpha, \beta)) \latticeVecHat{1} \\ \nonumber
     & = 2  \panelDimThree \sqrt{1 - \left(\frac{\cos(\alpha) + \cos(\beta)}{2 \sin(\foldAngle)}\right)^2} \times \\ \nonumber & \, \, \sqrt{1 - \left( \frac{\cos(\alpha) - \cos(\beta)}{2 \cos(\foldAngle) \sqrt{1 - \left[ \frac{\cos(\alpha) + \cos(\beta)}{2 \sin(\foldAngle)} \right]^2}} \right)^2} \latticeVecHat{1}
\end{align}
With this information, all of the vertices of the unit cell may be defined by transporting these vectors along one another, and the vertices of the infinite sheet may be defined from the unit cell by translation along the lattice vectors.

\section{Coarse formalism for origami deformations} \label{app:coarseModes}

Here we present a formalism to describe the coarse motion of an origami sheet, patterned with a repeating microstructure of folds. The Morph patterns constitute a representative set of examples, and the methods may be readily generalized to other fold patterns, such as those explored in Ref.~\cite{mcinerney2022discrete}. 

To describe the coarse macroscopic shape changes of an origami sheet, we obscure our view of the fine details of deformation within each unit cell, and instead focus on how each unit cell is positioned in space relative to its neighbors. This is accomplished using the smoothed midplane of the sheet, which intersects with the geometric center of each unit cell. In the case of the Morph, the unit cell geometry is represented by four vertices contained therein (see color-coded dots in main text Fig.~1a,b). In this work we choose the unit cell center as the geometric average of the four vertex positions.

The smooth sheet which the unit cell centers approximate, taken as a differentiable manifold, is governed by familiar quantities from differential geometry. For instance, it is well-known from the theorem of Bonnet~\cite{toponogov2006differential} that knowledge of the first and second fundamental forms is completely sufficient to describe the shape of this sheet, up to overall rotations and translations. These fundamental forms are defined from a parameterization of the smooth surface using two scalar parameters $(\sheetCoords)$.  Within this picture, the position of a point in the sheet is represented by a vector function $\posr(\sheetCoordOne, \sheetCoordTwo)$. 
The parameterization here is chosen to be the distance along each lattice direction in the undeformed lattice, such that the first fundamental form captures the metric of deformation, and may be used to quantify the strain induced in the sheet relative to the quasi-planar reference state. The tensor components of the first fundamental form are defined from the tangent vectors $\partial_i \posr$, where $\partial_i$ denotes the usual partial derivative with respect to the $i$th parameterization coordinate. Given the choice of parameterization, the tangent vectors then take the form $\partial_i \posr = \latticeVec{i} / |\init{\latticeVec{i}}|$, where $\latticeVec{i}$ is the lattice vector connecting the unit cell position to its neighbor in the $i$th direction  and $\init{\latticeVec{i}}$ is the same lattice vector prior to deformation, in the quasi-planar reference state.  The tensor components of the first fundamental form are then written 
\begin{equation}\label{eq:app_metricGeneral}
\metric_{ij} \equiv (\partial_i \posr) \cdot (\partial_j \posr) =\frac{ \latticeVec{i} \cdot \latticeVec{j}}{|\init{\latticeVec{i}}| |\init{\latticeVec{j}}|} \, .
\end{equation}
This tensor quantity may be used to define the Green-Lagrange nonlinear strain of the surface in its local plane via
\begin{equation}\label{eq:app_strainDef}
    \strain_{ij} = \frac{1}{2}(\metric_{ij} - \metric_{ij}^{(0)})
\end{equation}
in which $\metric_{ij}^{(0)}$ is the reference metric of the material, typically represented by the identity. Note that this expression for the metric relies on the assumption that these lattice vectors are much smaller than all other lengthscales of spatial variation of the midplane, and therefore that they approximate the sheet infinitesimals.

To capture the second fundamental form, we use these midplane surface tangent vectors to define the surface normal via: $\normal \equiv \frac{\latticeVec{1}\times\latticeVec{2}}{|\latticeVec{1}\times\latticeVec{2}|}$. From this, the components of the second fundamental form are obtained with the formula $\curvature_{ij} = -\partial_i \posr \cdot \partial_j \normal$. Note that due to the choice of sheet parameterization with units of length, that components of the second fundamental form have units of inverse length. In particular, the contraction with the inverse metric $\metric^{ik}\curvature_{kj}$ evaluated in a diagonal frame captures the \emph{principal curvatures}, which are the eigenvalues, and which reflect the inverse of the radii of curvature of the midplane of the sheet. Note that these curvatures vanish for the spatially-periodic rigid ground states of the sheets, as the sheet normal is then uniform.


\subsection{Triangulation and linear modes}\label{app:linearModesDefine}

As discussed in Sec.~\ref{app:foldPatterns}, a fold pattern in the Morph family has only a single rigid folding motion. However, each constitutive parallelogram panel is itself a thin elastic sheet. As explored in greater detail in Sec.~\ref{app:realParams}, thin sheets are qualitatively much softer to bending deformations than to stretching. In general cases of soft deformations of a thin sheet one might typically consider the highly dimensional space of isometries, in addition to the possibilities of singular structures such as d-cones and ridges. However, each of the parallelogram panels of an origami pattern such as the Morph are constrained to match to neighboring panels along straight edges, which greatly reduces the isometries available. As has become a standard protocol in origami modeling and analysis~\cite{Schenk2013, Wei2013, Filipov2017, Liu2017, liu2018highly, McInerney2020} the softest deformations of the panels maybe approximated by the addition of a crease across the diagonal of the quadrilateral face. With the addition of such ``panel bending'' crease lines, we describe the origami sheet as a triangulated mesh. This enables the use of ``bar-hinge'' type models, in which the complex deformation of an origami sheet to be captured in a reduced description where the degrees of freedom are the positions of the vertices.

This ``bar-hinge'' approximation has been very helpful in identifying the softest available motions for the large-scale origami sheet. Just as the very small comparative energy penalty of the folding at the major creases allowed us to identify the rigid folding motion, we may further choose to neglect the small energetic penalties that may arise from the panel bending. In this idealized case, the origami sheet resembles a truss, or a ball-spring network, with vertices connected by springs as in Fig.~\ref{fig:app_morph}c, and without any torsional energy penalties coming from triangular panel orientations. In this limit, one may use constraint counting and symmetry arguments to show that three infinitesimal zero energy modes of deformation must exist in the infinite system~\cite{McInerney2020}. One of these modes is the quasi-planar rigid folding motion established above, which extends nonlinearly. The other two only achieve linearly small values, and are known colloquially as the ``twist'' and ``bend'' modes, according to their qualitative effect on the sheet midplane. We here review and establish the nature of these three modes via their effect on midplane shape, as captured by the fundamental forms.

The first mode, referred to here as the ``planar'' mode, is parameterized by the folding angle $\foldAngle$, which may extend to nonlinear values on the order of $\pi$. At the coarse scale, the effect of this folding motion is captured in the first fundamental form via
\begin{equation}\label{eq:app_ribbonff1}
  \metric_{ij} \rightarrow \metric^*_{ij}(\foldAngle) \equiv \frac{\latticeVec{i}(\foldAngle) \cdot \latticeVec{j}(\foldAngle)}{|\latticeVec{i}(\init{\foldAngle})|  |\latticeVec{j}(\init{\foldAngle})|} \, ,
\end{equation}
where $\init{\foldAngle}$ is the folding angle in the chosen reference state. This equation is simply the special case of Eq.~\ref{eq:app_metricGeneral} in which the lattice vectors, and thereby the metric, are now determined by this folding angle. In the case of the Morph, the lattice vectors are orthogonal for all values of the folding, and this general formula therefore simplifies to
\begin{equation}\label{eq:app_ribbonff1_morph}
  \metric^*_{ij}(\foldAngle) \rightarrow \begin{bmatrix} \principalOne(\foldAngle)^2 & 0 \\ 0 & \principalTwo(\foldAngle)^2 \end{bmatrix} \, ,
\end{equation}
as shown in main text Eq.~\ref{eq:metric}, where
\begin{align}\label{eq:app_principalOneDefine}
\principalOne(\foldAngle) & = \frac{ |\latticeVec{1}(\foldAngle)| }{ |\latticeVec{1}(\init{\foldAngle})| } \\ \label{eq:app_principalTwoDefine}
\principalTwo(\foldAngle) & = \frac{ |\latticeVec{2}(\foldAngle)| }{ |\latticeVec{2}(\init{\foldAngle})| }
\end{align}
are the principal stretches induced by the planar mode and, for the Morph, $\latticeVec{1}, \latticeVec{2}$ are the functions defined above in Eqs.~\ref{eq:app_latticeVecMorphOne}\&\ref{eq:app_latticeVecMorphTwo}.

To understand the second and third modes, referred to here (as well as in other works) as the ``twist'' and ``bend'' modes, we must consider the second fundamental form $\curvature_{ij}$. As noted above, this tensor is symmetric, and therefore determined by three scalars, $\curvature_{11}$, $\curvature_{12}$ and $\curvature_{22}$. As noted in previous work, the twist and bend modes may be fruitfully described in terms of dihedral angles of the triangulated sheet, or, in terms of ``vertex'' and ``face'' amplitudes~\cite{mcinerney2022discrete}. However, in either description each mode may only  achieve linearly small values before they begin to cause sheet stretching. Further, as the only nonlinear mode available to the origami sheet $\foldAngle$, does not contribute to this curvature tensor, the curvature must also only extend to linear order in the detailed degrees of freedom. Therefore, the components of this tensor may be written in in a basis of three amplitudes  
\begin{equation}\label{eq:app_ribbonff2_morph}
    \curvature_{ij} \rightarrow \curvature^*_{ij}(\foldAngle, \twist, \bend) = \twist \hat{\twist}_{ij}(\foldAngle) + \bend \hat{\bend}_{ij}(\foldAngle) + \thirdCurvature \hat{\curvature}^{(3)}_{ij}(\foldAngle)
\end{equation}
chosen so that the first two terms capture the twist and bend fields, while the third term captures the component of curvature that does not arise from bend or twist and therefore we will have $\thirdCurvature=0$ in our hyperribbon theory. Our task is therefore to determine the nature of the curvature subspace we are restricted to, defined by the basis $\hat{\twist}(\foldAngle)$ and $\hat{\bend}(\foldAngle)$. 

To determine this subspace, we first allocate the off diagonal component of the curvature to the twist mode by setting
\begin{equation}\label{eq:app_twistHat}
   \hat{\twist}(\foldAngle) = \begin{bmatrix} 0 & 1 \\ 1 & 0 \end{bmatrix} \, .
\end{equation}
This indicates that the twist mode will generate no curvature along the lattice vector directions, in accordance with prior twist mode descriptions~\cite{Schenk2013, mcinerney2022discrete}. Note that this definition removes any dependence of $\hat{\twist}$ on the folding angle $\foldAngle$. 

Within this definition of the twist, the generic form for the bend mode curvature must take the form
\begin{equation}
  \hat{\bend} = \begin{bmatrix} \bend_1(\foldAngle) & 0 \\ 0 & \bend_2(\foldAngle) \end{bmatrix} \, .
\end{equation}
Rather than requiring additional computation of the unit cell structure, the functions $\bend_1$ and $\bend_2$ may be related to the principal stretches $\principalOne$ and $\principalTwo$. Incrementation of the bend mode generates (and defines) the out-of-plane Poisson's ratio
\begin{equation}
  \poissonOut \equiv -\frac{\delta\curvature_{22}}{\delta\curvature_{11}} = -\frac{\bend_2(\foldAngle)}{\bend_1(\foldAngle)}
\end{equation}
while incrementation of the planar mode defines the in-plane Poisson's ratio
\begin{equation}
  \poissonIn \equiv -\frac{\delta\metric_{22}}{\delta\metric_{11}} \, .
\end{equation}
Knowing the form of the metric, and its increment, we may write this in terms of the principal stretches
\begin{equation}
  \poissonIn(\foldAngle) = -\frac{\principalTwo(\foldAngle) \principalTwo'(\foldAngle)}{\principalOne(\foldAngle) \principalOne'(\foldAngle)}
\end{equation}
where prime denotes derivative with respect to the argument $\foldAngle$. It has been well-established that folding patterns in this category have $\poissonIn = -\poissonOut$~\cite{Schenk2013, Wei2013, Pratapa2019, mcinerney2022discrete, Nassar2022} and we may use this relation to write
\begin{align}\label{eq:app_bendOneRelation}
  \bend_1(\foldAngle) = & \principalOne(\foldAngle) \principalOne'(\foldAngle) \\ \label{eq:app_bendTwoRelation}
  \bend_2(\foldAngle) = & -\principalTwo(\foldAngle) \principalTwo'(\foldAngle)
\end{align}
without any loss of generality. We are free to choose to put the negative sign on the $\bend_2$ term as this choice may be absorbed into our definition of the bend mode amplitude itself, along with any other fold-dependent multiplicative factor applied to both $\bend_1$ and $\bend_2$ equally. Again, due to the symmetry of the origami pattern, this bend mode does not contribute to the first fundamental form to first order. Despite matching qualitatively with existing literature, these particular amplitude definitions for the twist and bend modes are unique to this work, having been chosen for analytic convenience in our coarse description.


\subsection{Spatially varying planar, twist, and bend modes}\label{app:spatialSoftModes}


Following the central statement of the main text regarding spatially varying generalizations of these local low-energy modes, we promote these scalar mode parameters to fields $\{\foldAngle, \bend, \twist\} \rightarrow \{\foldAngle(\sheetCoords), \bend(\sheetCoords), \twist(\sheetCoords) \}$ on the reference space (i.e. surface parameterization). One may think of these fields, and their gradients, as constituting a recipe which determines the change in crease folding and panel bending in each unit cell with respect to the reference state. It therefore follows that the field modes constitute a recipe to determine the fundamental forms across the sheet. In principal, all of the field values as well as their gradients may contribute to determine the fundamental forms. We are interested in deformations where the gradients are small, as are the dimensionless twist and bend modes themselves $\panelDimOneTwo\,\bend(\sheetCoords), \panelDimOneTwo\,\twist(\sheetCoords)$. To lowest order approximation in these small quantities, we are therefore concerned with the influence  of the nonlinear planar mode $\foldAngle(\sheetCoords)$ as well as the influence of the linear effects of the planar mode gradients $\nabla_{\sheetCoords} \foldAngle$ and the twist and bend. However, to the symmetry of the Morph folding pattern upon mirroring in the $\sheetCoordOne$ and $\sheetCoordTwo$ directions, the gradients of the planar mode may not contribute to the first fundamental form. We are therefore left with the simple generalization of the above relations for the effect of the uniform modes on the fundamental forms
\begin{align} \label{eq:app_metricFieldAssumption}
  \metric_{ij}(\sheetCoords) & \rightarrow \metric^*_{ij}(\foldAngle(\sheetCoords)) \\ \label{eq:app_curvatureFieldAssumption}
  \curvature_{ij}(\sheetCoords) & \rightarrow \curvature^*_{ij}(\foldAngle(\sheetCoords),  \twist(\sheetCoords), \bend(\sheetCoords)) \, .
\end{align}
Corrections to these expressions may arise at higher order, proportional to second gradients of the folding angle $\foldAngle$, as well as first gradients of the twist and bend. 

With Eqs~\ref{eq:app_metricFieldAssumption}\&\ref{eq:app_curvatureFieldAssumption}, we have determined that the possible 6 local shape degrees of freedom $\metric_{11}, \metric_{12}, \metric_{22}, \curvature_{11}, \curvature_{12}, \curvature_{22}$ of the origami midplane will be determined entirely by three parameters (the local soft modes), and therefore must be restricted to a 3 dimensional subspace of this possible 6 dimensional space. We may characterize this 3 dimensional subspace by first considering the nonlinear mode $\foldAngle(\sheetCoords)$. In the absence of the other two mode-fields, the origami sheet must lie on a 1 dimensional nonlinear manifold contained within the 3 dimensional shape space. Around each of these points, the possible inclusion of the twist and bend fleshes out this one dimensional manifold to become locally 3 dimensional. This structure in the 6 dimensional shape space which is locally long in one direction, short in another (2) directions, and flat in the additional directions is reminiscent of a three dimensional ribbon structure, and we therefore refer to this as the hyperribbon.



\subsection{Compatibility and the Gauss-Codazzi equations}\label{app:gaussCodazzi}


As identified above, the spatially varying planar, bend, and twist mode fields are expected to capture the response of the origami sheet. One might imagine that any spatially varying pattern of these modes will correspond to a candidate soft response. However, as also mentioned above, we may think of these modes as a recipe determining the changes in the shape of the sheet midplane. It is well established that in the continuum picture of such a sheet, the strains and curvatures of the midplane must obey compatibility relations, or else when we attempt to follow the recipe from the modes to determine shape of the deformed sheet we will encounter geometric inconsistencies. The problem of closure must be addressed to guarantee that a particular mode pattern is geometrically valid, and therefore corresponds to a possible continuous sheet configuration in three dimensions. 

While approaches to the compatibility of the sheet may be analyzed using loop closure relations, it is known that this question is readily addressed by the Gauss-Codazzi-Mainardi-Peterson equations. That the first and second fundamental forms must satisfy these equations is a requirement for compatibility. Compatibility is required to ensure that the sheet may be embedded in 3-dimensional space (i.e. that such a surface exists having the corresponding pattern of fundamental forms). Further, according to the theorem of Bonnet~\cite{toponogov2006differential}, this constitutes a sufficient set of conditions to guarantee compatibility and therefore also guarantees a corresponding low-energy configuration of the origami sheet.

The Gauss-Codazzi equations are captured in compact form in two parts. First by the Gauss equation, which is a single scalar equation:
\begin{equation}\label{eq:app_gaussDef}
  \scalarCurvature = 2\gaussCurvature
\end{equation}
where
\begin{equation}\label{eq:app_gaussCurvatureDef}
    \gaussCurvature \equiv  \text{det}[\curvature^i_j] = 2 \text{det}[\metric^{ik} \curvature_{kj}]
\end{equation}
is the Gaussian curvature,
\begin{equation}\label{eq:app_scalarCurvatureDef}
  \scalarCurvature \equiv \metric^{ik}\riemann^j_{ijk}
\end{equation}
is the (Ricci) scalar curvature defined in terms of the the inverse metric $\metric^{ij}$ (satisfying $\metric^{ij}\metric_{jk} = \delta^i_k$) and the Riemann curvature tensor capturing the intrinsic curvature of the sheet
\begin{equation}\label{eq:app_riemannDef}
  \riemann^l_{ijk} \equiv \partial_j \christo^l_{ik} - \partial_k \christo^l_{ij} + \christo^l_{js} \christo^s_{ik} - \christo^l_{ks}\christo^s_{ij}
\end{equation}
which is in turn defined using the Christoffel symbol
\begin{equation}\label{eq:app_christoffelDef}
  \christo^l_{ij} \equiv \frac{1}{2}\metric^{lk} \left( \partial_j\metric_{ki} + \partial_i\metric_{kj} - \partial_k\metric_{ij} \right) \, .
\end{equation}

Complementary to the Gauss equation are the Codazzi equations, which are captured by the set of first order PDEs
\begin{align}\label{eq:app_codazziOneDef}
  \levi^{ij}D_i \curvature_{jk} = 0
\end{align}
which must hold for $k = \{1, 2\}$. Here $D_i$ is the covariant derivative defined by
\begin{equation}\label{eq:app_covariant}
  D_i \curvature_{jk} \equiv \partial_i \curvature_{jk} - \christo^l_{ij} \curvature_{lk} - \christo^l_{ik}\curvature_{lj} \, ,
\end{equation}
and $\levi^{ij}$ is the totally antisymmetric tensor defined by component as $\levi^{11} = \levi^{22} = 0$ and $\levi^{12} = -\levi^{21} = 1$.

These equations, for a general surface, impose three PDE constraints on the six otherwise independent fields $\metric_{11}(\sheetCoords), \metric_{12}(\sheetCoords), \metric_{22}(\sheetCoords), \curvature_{11}(\sheetCoords), \curvature_{12}(\sheetCoords), \curvature_{22}(\sheetCoords)$. However, for the soft origami deformations, the six fundamental form fields are determined by the three mode fields  $\{\foldAngle, \bend, \twist\} \rightarrow \{\foldAngle(\sheetCoords), \bend(\sheetCoords), \twist(\sheetCoords) \}$. We therefore rewrite and simplify the GC equations by inserting the fundamental forms from Eqs.~\ref{eq:app_ribbonff1_morph}\&\ref{eq:app_ribbonff2_morph}. This process is tedious but straightforward, the results detailed below. 

For the Eq.~\ref{eq:app_gaussDef} above, we first compute the Gaussian curvature $\gaussCurvature = \text{det}[\metric^{ik}\curvature_{kj}]$ which is the product of the sheet principal curvatures. This reduces to 
\begin{equation}\label{eq:app_gaussCurvatureFinal}
  \gaussCurvature = \frac{1}{[\principalOne(\foldAngle) \principalTwo(\foldAngle)]^2} \left(\bend^2 \bend_1(\foldAngle) \bend_2(\foldAngle) -  \twist^2 \right) \, .
\end{equation}
Next, the scalar curvature, which is the most tedious, may be expanded and simplified before inserting our particular forms
\begin{align}
  \scalarCurvature = & \metric^{ik} \left[ \partial_j \christo^j_{ik} - \partial_k\christo^j_{ij} + \christo^j_{js}\christo^s_{ik} - \christo^j_{ks}\christo^s_{ij} \right] \\ \nonumber
    = & \frac{1}{2} \metric^{ik} \Big\{ (2 \partial_i\metric_{kl} - \partial_l \metric_{ik})\partial_j \metric^{jl} - \partial_i\metric_{jl}\partial_k\metric^{jl} \\ \nonumber 
    & \qquad + 2 \metric^{jl} \left(  \partial_i \partial_j \metric_{kl} - \partial_i \partial_k \metric_{jl}\right) \\ \nonumber 
    & \qquad + \frac{\metric^{jm} \metric^{ln}}{2} \Big[ 2 \partial_l \metric_{mj} \partial_i\metric_{nk} - \partial_l \metric_{mj} \partial_n \metric_{ik} \\ \nonumber
    & \qquad - 2 \partial_l \metric_{mk} \partial_i\metric_{nj} + \partial_l \metric_{mk}\partial_n\metric_{ij}\Big]\Big\}\, .
\end{align}
After plugging in, this reduces to 
\begin{align}\nonumber
    \scalarCurvature & = \frac{2}{\principalOne \principalTwo} \Bigg[ \frac{\principalOne' \principalTwo'(\partial_2 \foldAngle)^2}{\principalTwo^2} + \frac{\principalOne' \principalTwo'(\partial_1 \foldAngle)^2}{\principalOne^2} \\ \label{eq:app_scalarCurvatureFinal} & -\frac{\principalTwo'' (\partial_1\foldAngle)^2}{\principalOne} - \frac{\principalOne'' (\partial_2\foldAngle)^2}{\principalTwo}  - \frac{\principalTwo' \partial_1 \foldAngle}{\principalOne}  - \frac{\principalOne' \partial_2 \foldAngle}{\principalTwo}   \Bigg]
\end{align}
 where, as a reminder, $\principalOne$ and $\principalTwo$ and their derivatives (primes) are both functions of the mechanism parameter field $\foldAngle$ even though this dependence is suppressed here for ease of presentation. The Gauss equation then may be assembled as a condition on the fields $\foldAngle, \bend, \twist$ by placing these relations Eq.~\ref{eq:app_scalarCurvatureFinal}\&\ref{eq:app_gaussCurvatureFinal} into Eq.~\ref{eq:app_gaussDef}, leading to main text Eq.~\ref{eq:gauss}. 

Similar to the Gauss equation, the Codazzi equations reduce to 
\begin{align} \nonumber
   0  = & \left( \partial_1 \twist - \principalOne \principalOne' \partial_2 \bend - \bend (\principalOne\principalOne'' + \principalOne'\principalOne')\partial_2 \foldAngle \right) \\ \label{eq:app_codazziOneFinal}
   & + \left( \frac{\principalTwo'}{\principalTwo} - \frac{\principalOne'}{\principalOne} \right) \twist \partial_1 \foldAngle + \left( \frac{ \principalOne'}{\principalOne} + \frac{ \principalTwo'}{\principalTwo}  \right) \bend \principalOne \principalOne' \partial_2 \foldAngle
\end{align}
and
\begin{align} \nonumber
   0  = & - \left( \partial_2 \twist + \principalTwo \principalTwo' \partial_1 \bend + \bend (\principalTwo\principalTwo'' + \principalTwo'\principalTwo')\partial_1 \foldAngle \right) \\ \label{eq:app_codazziTwoFinal}
   & + \left( \frac{\principalTwo'}{\principalTwo} - \frac{\principalOne'}{\principalOne} \right) \twist \partial_2 \foldAngle + \left( \frac{ \principalOne'}{\principalOne} - \frac{ \principalTwo'}{\principalTwo}  \right) \bend \principalTwo \principalTwo' \partial_1 \foldAngle \, .
\end{align}
Combined with the Gauss equation, this imposes three PDE relations on the three soft mode fields. This balance of equations and fields indicates that the solution in the interior will be determined by the boundary conditions. Specifically, the geometry of the soft mode will be determined by boundary conditions on the fields $\foldAngle, \bend, \twist$ Nonlinearly these may be restricted by the lack of known existence and uniqueness theorems, yet the number of possible solutions in the interior are expected to be proportional to the boundary. The soft modes are therefore infinite in dimension, yet subextensive. 


\subsection{Linear Deformations}\label{app:linearAnalytics}

The equations of compatibility determined in the previous section govern the soft response of the origami metamaterial: each solution to these equations guarantees a corresponding soft pattern of response which may arise under loading. Therefore our task is to locate the space of solutions to these reduced Gauss and Codazzi equations. For nonlinear strains, analytic solutions are not obvious. However, linearizing these equations in the perturbative modes $\twist(\sheetCoords), \bend(\sheetCoords), \delta\foldAngle(\sheetCoords)$ allows closed-form solutions to be obtained. Here, $ \delta\foldAngle(\sheetCoords) \equiv \foldAngle(\sheetCoords) - \init{\foldAngle}$ implicitly defines the quasi-planar state of folding (i.e. mechanism reference state) we are deforming from. \mdc{Such analysis makes clear the spatial character of these soft modes, as well as the dualilty indicated in the main text.}

Using $\principalOne(\init{\foldAngle}) = \principalTwo(\init{\foldAngle}) = 1$, the linearized equations appear as
\begin{align}\label{eq:app_gaussLinear}
    0 & = (\partial^2_1 - \frac{1}{\poissonIn} \partial^2_2) \delta \foldAngle \\ \label{eq:app_codazziOneLinear} 
    0 & = \partial_1 \twist - \principalOne' \partial_2 \bend \\ \label{eq:app_codazziTwoLinear}
    0 & = \partial_2 \twist + \principalTwo' \partial_1 \bend
\end{align}
where
\begin{equation}\label{eq:app_poissonLinear}
\poissonIn
     = \frac{-\principalTwo'}{\principalOne'}
\end{equation}
is the in-plane linear Poisson's ratio of the planar folding mode around this folding state $\init{\foldAngle}$. These equations may be rewritten in a suggestive form
\begin{align}\label{eq:app_gaussWform}
    0 & = \partial_\ww \partial_{\bar{\ww}} \delta \foldAngle \\ \label{eq:app_codazziOneWform}
    0 & = \partial_{\bar{\ww}} \curvatureVar \\ \label{eq:app_codazziTwoWform}
    0 & = \partial_\ww \bar{\curvatureVar}
\end{align}
where we have introduced new variables for the surface parametrization
\begin{align}\label{eq:app_w}
    \ww & = \sheetCoordOne + \frac{1}{\gamgam}\sheetCoordTwo \\ \label{eq:app_wBar}
    \bar{\ww} & = \sheetCoordOne - \frac{1}{\gamgam}\sheetCoordTwo
\end{align}
and another set of new variables capturing the soft curvature fields
\begin{align}\label{eq:app_curvature}
    \curvatureVar & = \twist - \gamgam  \principalTwo' \bend \\ \label{eq:app_curvatureBar}
    \bar{\curvatureVar} & = \twist + \gamgam  \principalTwo' \bend  \, .
\end{align}
Here, $\gamgam = \sqrt{\frac{1}{\poissonIn}}$ is defined for convenience and the derivatives
\begin{align}
    \partial_\ww & = \frac{1}{2} \left( \partial_1 + \gamgam \partial_2 \right) \\ 
    \partial_{\bar{\ww}} & = \frac{1}{2} \left( \partial_1 - \gamgam \partial_2 \right)
\end{align}
are implied by the requirements that $\partial_\ww \ww =  \partial_{\bar{\ww}} \bar{\ww} = 1$ and $\partial_\ww \bar{\ww} = \partial_{\bar{\ww}} \ww = 0$. This compatibility theory has notably decoupled the fields that generate curvature $\twist, \bend$ from the planar mode $\foldAngle$, into two compatibility problems that may be solved independently. In the linear limit, these therefore divide into the ``planar'' soft modes (those involving $\foldAngle$) which do not generate curvature and the ``non-planar'' modes involving $\twist$ and $\bend$ which do not influence the strain on the surface. 

The choices of transformed variables in Eqs.~\ref{eq:app_w},\ref{eq:app_wBar},\ref{eq:app_curvature}\&\ref{eq:app_curvatureBar} are based largely on that introduced in~\cite{Czajkowski2022-2}, in which a related formalism enabled general closed-form solutions to be written for the planar unimode problem. Similarly, to solve for the non-planar soft modes, we note that all solutions to  Eqs.~\ref{eq:app_codazziOneWform}\&\ref{eq:app_codazziTwoWform} must be written as $\curvatureVar(\ww, \bar{\ww}) \rightarrow \curvatureVar(\ww)$ and $\bar{\curvatureVar}(\ww, \bar{\ww}) \rightarrow \bar{\curvatureVar}(\bar{\ww})$. However, these must also be required to generate real-valued fields for $\twist = \frac{1}{2} (\bar{\curvatureVar} + \curvatureVar)$ and $\bend = \frac{1}{2 \principalTwo' \gamgam} (\bar{\curvatureVar} - \curvatureVar)$. Tuning the origami metamaterial from anauxetic $\poissonIn>0$ to auxetic $\poissonIn<0$ changes the spatial coordinates $\ww, \bar{\ww}$ from pure real to complex. For the auxetic origami sheet, we require that $\curvatureVar(\ww)^* = \bar{\curvatureVar}(\bar{\ww})$, while for the anauxetic examples, these functions are simply required to be real-valued themselves. Here, $^*$ is the conventional complex conjugation.  

The utility of this analytic result is displayed in the case of origami sheets with the topology of the disc. In the auxetic case all solutions may be assembled from analytic expansion $\curvatureVar(\ww) = \sum_n C_n \ww^n$ with arbitrarily chosen complex coefficients, which then implies the expansion $\bar{\curvatureVar}(\ww) = \sum_n D_n \bar{\ww}^n$, with $D_n = C_n^*$. Similarly, the solutions for the anauxetic case are captured by the same sums, but with the coefficients required to be real-valued. 

 Having solved the compatibility of the twist and bend modes, we turn our attention to the planar mode. This particular question of planar motion compatibility is again a previously solved problem. As shown in~\cite{Czajkowski2022-2} the compatibility may also be usefully written 
\begin{align}
    \partial_{\bar{\ww}} \strainVar & = 0 \\
    \partial_{\ww} \bar{\strainVar} & = 0
\end{align}
in terms of the $\ww, \bar{\ww}$ coordinates. Here, we have defined variables capturing linear strain
\begin{align}
    \strainVar & = \aA \mechStrain - \gamgam \coarseRot \\ 
    \bar{\strainVar} & = \aA \mechStrain + \gamgam \coarseRot
\end{align}
where $\mechStrain$ is the (field) amplitude of strain arising from the linear folding mechanism $\delta \foldAngle$ and $\aA = \sqrt{2/(1 + \poissonIn^2)}$ is a mechanism strain normalization, and $\coarseRot$ is the coarse rotation of each continuum sheet element in the plane. More explicitly, the mechanism strain field is defined so that the observed unsymmetrized coarse strain $\unsymStrain_{ij} = \partial_j u_i $ from planar coarse displacements $u_1, u_2$ will appear as
\begin{equation}
    \ttens{\strain} = \mechStrain \aA \begin{bmatrix} 1 & 0 \\ 0 & -\poissonIn \end{bmatrix} + \coarseRot \begin{bmatrix} 0 & -1 \\ 1 & 0 \end{bmatrix} \, .
\end{equation}
Finally, $\mechStrain = \delta\foldAngle \frac{\principalOne'}{\aA}$ relates these scalar variables of planar strain.

The upshot of this additional formalism for linear deformations is that the duality between the planar deformations and non-planar deformations is now made explicit. Solutions for the curvature-inducing fields $\curvatureVar, \bar{\curvatureVar}$ may readily be converted into solutions for the planar deformation fields $\strainVar, \bar{\strainVar}$ and vice versa. This is accomplished with an arbitrary multiplicative factor $\unitConversionFactor$ with units of inverse length so that the nonplanar mode corresponds to the planar mode via $ (\curvatureVar, \bar{\curvatureVar}) \rightarrow \unitConversionFactor*(\strainVar, \bar{\strainVar})$. Of course, this may also be accomplished using the soft mode fields $\mechStrain, \coarseRot, \twist, \bend$ directly. Furthermore, elements of each of the dual soft mode spaces may be superimposed to span the soft deformations available to the origami sheet.

It is straightforward to show that each field $\mechStrain, \coarseRot, \twist, \bend$ each individually satisfies the same second order PDE: $\partial_{\bar{\ww}}\partial_\ww \foldAngle = \partial_{\bar{\ww}}\partial_\ww \coarseRot = \partial_{\bar{\ww}}\partial_\ww \bend = \partial_{\bar{\ww}}\partial_\ww \twist = 0$. To make this a bit more tangible, note that the second derivative $\partial_{\bar{\ww}} \partial_{\ww}$ reduces to $\frac{1}{4}(\partial^2_1 - \frac{1}{\poissonIn} \partial^2_2)$, which is a distorted Laplacian. Despite neglecting the connection between $\coarseRot$ and $\mechStrain$ and that between $\twist$ and $\bend$, these relations capture the (identical) spatial patterns available to each mode. Each mode may be written as the sum of two analytic functions i.e. $\mechStrain = \frac{1}{2 \aA} ( \strainVar(\ww) + \bar{\strainVar} (\bar{\ww}) ) $. Again, these functions may be independent real analytic functions in the anauxetic case, and must be complex conjugates in the auxetic case. As explored in more depth in~\cite{Czajkowski2022-2}, this means that the auxetic patterns will appear as boundary modes, which decay exponentially into the bulk, while the anauxetic modes will appear as periodic functions composed of plane waves. These characteristic behaviors are separated by an exceptional point in the Poisson's ratio at $\poissonIn=0$, where the coordinate transformations $\ww$ and $\bar{\ww}$ are identical and the mode becomes uniaxial.




\section{Simulation of Origami Sheets} \label{app:simulations}

Here we describe the process used to generate numerically force-balanced states of origami sheets under point loading conditions.

\subsection{Numerical investigations in a reduced model using the MERLIN software}\label{sec:merlin}

To investigate the bulk force-balanced states available to the periodic fold patterns in the Morph family, we employ the MERLIN2 software~\cite{Liu2017, Liu2018} developed by the Paulino research group at Georgia Tech. This software, which runs in MATLAB, treats the origami sheet using a bar-hinge model with the vertex positions (orange points in Fig.~\ref{fig:app_morph}c) as the degrees of freedom. To generate data for this manuscript, we employ the N4B5 triangulation scheme, in which a crease line is added to each panel along the shorter of the diagonals as described below. This software has been highly accurate in describing the deformation of real experimental origami sheets~\cite{Liu2017}. As described in the main text, deformations of these vertices incur energy penalties of three categories: bar stretching, crease folding, and panel bending, captured by the terms
\begin{equation}
    E = \energyStretch + \energyFold + \energyBend
\end{equation}
respectively.

Bar stretching penalizes changes in distance between any pair of nodes connected by an edge such as the green and blue lines in Fig.~\ref{fig:app_morph}c. An example of such a bond connects node $\bondIndexNodeOne$ to $\bondIndexNodeTwo$. The vector $\vvec{r}_{\bondIndex}$ connects these nodes in space and based on this the stretching energy penalty is 
\begin{equation}
   \energyStretch = \sum_{\text{bonds} \langle\bondIndex\rangle} \frac{1}{2} \barArea \bondLength_{\bondIndex} W(\frac{(|\vvec{r}_{\bondIndex}| - \bondLength_{\bondIndex})}{\bondLength_{\bondIndex}})
\end{equation}
%
where $\barArea$ is a cross-sectional area of the bar element, and $\bondLength_{\bondIndex}$ is the distance between the nodes in the undeformed state (i.e. length of the bar element). The bar energy density function
\begin{equation}
    W(x) = \modStretch \left( \sqrt{2x+1} - \frac{(2x+1)^{5/2}}{5} \right) 
\end{equation}
captures an Ogden nonlinear constitutive model~\cite{Liu2017}.  This nonetheless resembles a system of conventional linear Hookean springs
\begin{equation}\label{eq:app_stretchingEnergy}
    \energyStretch \sim \sum_{\text{bonds} \langle\bondIndex\rangle} \frac{\barArea \modStretch}{2 \bondLength_{\bondIndex}} (|\vvec{r}_{\bondIndex}| - \bondLength_{\bondIndex})^2
\end{equation}
for small strains of the bar elements. Even for the nonlinear macroscopic deformations explored in main text Fig.~\ref{fig:main1}~and~\ref{fig:main2}, the bar elements are only strained by linearly small amounts. Note that $\modStretch$ has the units of a compression modulus or Young's modulus ($J/m^3$). 

The panel bending is intended to capture the effect indicated in the name. Each parallelogram panel is itself a thin plate, which is therefore susceptible to isometric deformations. These deformations are approximated by the addition of a crease along the shorter of the two diagonals of the parallelogram. The addition of this crease defines two triangular panels of the parallelogram, and the panel bending energy penalizes the difference between orientation of the normal direction of each panel, as indicated by the dihedral angle $\dihedral_\panelIndex$ in Fig.~\ref{fig:app_morph}c. The energy penalty is computed from these dihedral angles of panel bending via
\begin{equation}\label{eq:app_energyBend}
    \energyBend = \sum_{\text{panels}- \panelIndex} \bondLength_\panelIndex \modBend (\dihedral_\panelIndex - \pi)^2 \, ,
\end{equation}
where the sum is taken over panels $\panelIndex$, $\bondLength_\panelIndex$ is the rest length of the panel diagonal crease line and $\modBend$ is the modulus of panel bending, which has the units of line tension. 

This crease line across the panel diagonal is also included as a panel stretching spring in Eq.~\ref{eq:app_stretchingEnergy}. This is essential to prevent the simulations from accessing unrealistic zero energy deformations which would strain the otherwise stiff parallelogram panels they aim to capture.

The energy of crease folding is captured in a similar manner to the panel bending, but with a distinct rotational stiffness and the dihedral angles having a rest position distinct from $\pi$. To calculate this energy requires knowledge of each of the rest state dihedral angles $\dihedral_\creaseIndex^0$ at each crease $\creaseIndex$ (i.e. given the initial quasi-planar state of the morph, what is the angle of the crease $\creaseIndex$?). From this the energy is calculated via
\begin{equation}\label{eq:app_energyCrease}
    \energyFold = \sum_{\text{creases} \creaseIndex} \bondLength_\creaseIndex \modFold (\dihedral_\creaseIndex - \dihedral_\creaseIndex^0)^2 \, ,
\end{equation}
where the sum is taken over all the (bulk) crease segments of the folding pattern, and $\modFold$ is the modulus of crease folding.

\subsection{Applied loading and constitutive parameters} \label{app:realParams}

The main text data is categorized into two types of constitutive models, designated the ``rigid'' versus the ``realistic'' origami sheets. These designations determine the combinations of the sheet constitutive parameters $\modStretch, \modBend, \modFold, A$ used in the simulations (as defined in Sec.~\ref{sec:merlin}). The primary difference between these is the presence of a finite stiffness against folding at the creases in the ``realistic'' version, which is absent in the ``rigid'' origami. They are therefore named according to their intended interpretation, where ``realistic'' emulates a folded sheet of material, while ``rigid'' emulates origami assembled from rigid panels connected by comparatively energy-free hinges.  For all the simulation data herein, lengths are measured against the uniform crease length $\panelDimOne = \panelDimTwo = \panelDimThree = 1$.

To obtain the so-named ``realistic'' origami sheet configurations displayed in Figs.~\ref{fig:main1}\&\ref{fig:main2}, we employ the ``auto'' mode within the MERLIN2 code. Here, the constitutive parameters for the bar-hinge model are specified indirectly, via the specification of the Poisson's ratio $\poissonMater$ and Youngs modulus $Y$ of the thin sheet, as well as a quantity which controls the folding stiffness of the panel diagonals relative to the folded crease lines $\ellRatio = \frac{\bendingModChris}{\modFold}$. The factor
\begin{equation}\label{eq:modBendChris}
\bendingModChris = \frac{Y \sheetThickness^3}{12(1 - \poissonMater^2)}
\end{equation}
is the bending modulus of the constitutive thin sheet, which is determined by $Y$ and $\poissonMater$, and in turn determines the stiffness of the panel bending and crease line bending. 
We assign these parameters to values which are appropriate for conventional thin sheets, choosing $\poissonMater = 0.3$, $\sheetThickness=0.01$, $\ellRatio = 20$. The two origami patterns simulated in main text Fig.~\ref{fig:main1} have crease geometry set by $\beta =1$, $\alpha = \pi - \beta$. The folding is set to $\gamma=0.01$ for the top imitation of an ``uncreased sheet'' in Fig.~\ref{fig:main1}, which appears to be a sufficiently small folding to remove the exotic effects of the crease pattern, while the lower folded portion is set to a folding angle $\gamma=0.5$.  Recall the precise definition of $\gamma$ presented in App.~\ref{app:foldPatterns}. The loading here is chosen to be incompatible with a cylindrical configuration, expected for the isometries of least bending and therefore least (small) energy. Then for main text Fig.~\ref{fig:main2}, the identical origami sheet is simulated as the folded example from Fig.~\ref{fig:main1}, but with different (arbitrary) displacement loading conditions applied to the four corners of the sheet. 
For all of the simulations herein, the energy scale is determined by choosing the Young's modulus $Y = 1$.


For the ``rigid'' data, we utilize the ``manual'' setting for the MERLIN2 simulation framework, allowing us to more directly set the parameters $\modFold, \modBend, \modStretch$ and the cross-sectional area for the bar elements $\barArea$. In this ``rigid'' data, the defining feature is that the folding modulus of the major creases $\modFold$ is set to zero, thereby realizing the ideal mechanism limit, in which the planar mode costs zero energy. The scale of energy is determined by the chosen panel bending modulus $\modBend = 1$. Aside from these choices, and aside from the parameters we explicitly vary as indicated, we 
use representative values for a sheet of paper, or similar material, with thickness $\sheetThickness$. For a common piece of origami folded from copy paper, the thickness (relative to the crease length) may be $\sim 0.01$. 

The remaining undetermined parameters required for these simulations in the  manual setting are $\modStretch, \modBend$ and $\barArea$. To provide context for these and estimate reasonable values, we first consider the bar elements, which capture the stretching of the sheet in its own plane. In this case the cross sectional area of a bar-element approximation of such phenomena will be approximately $\barArea \sim \sheetThickness l_{\nu}$, and each ``rigid'' simulation is hence run with $\barArea=0.01$. We may roughly approximate $\modStretch$ with the Young's modulus $Y$. In contrast, the stiffness against bending of a panel will scale as $\modBend \sim \frac{\sheetThickness^3 Y}{l_{\nu}}$, where the denominator contains the linear dimension of the curved portion of the panel, which we presume to be approximated by the edge length. Then the ratio between the stretching and panel bending moduli will be of the order $\frac{\modStretch}{\modBend} \sim \frac{t^3}{l_{\nu}} \sim 10^6$ in the natural length units of the system (chosen to be $l_{\nu}$). Or, connecting to the dimensionless parameter in Fig.~\ref{fig:main5}, $\frac{\modStretch}{a \modBend} \sim 10^6$. Therefore, with the the length $\panelDimOne$ set to unity, the default value of $\modStretch$ in simulations will be $10^6$. This dramatic discrepancy between the energy cost of deformations which stretch the constitutive material versus those which simply bend (i.e. fold) reflects the isometric limit.


The dataset displayed in main text Fig.~\ref{fig:main3} is obtained using simulations of origami sheets which extend by 20 unit cells in each lattice direction. As suggested by the above analysis, the stretching modulus is set to $\modStretch = 10^6$ for these ``rigid'' simulations. These are run with reference configuration geometry defined by the parameters in Table.~\ref{tab:nonlinearRunsConfig}. Each is run with an identical displacement loading applied to the vertices at the four corners of the quasi rectangular origami sheet. In this loading the vertices are displaced according to the vectors $\{ \vvec{d}_i\} = \{ \{ 0.8, 2.4, -2.0 \}, \{ 1.4, -0.8, 2.4\}, \{ -2.8, 2.4, -3.4 \}, \{ 0.6, -4.0, 3.0 \} \}$ (listed counter-clockwise from the bottom-left in the coordinate system defined by the orthonormal vectors of the initial configuration $\latticeVecHat{1}, \latticeVecHat{2}, \normal$ respectively). The simulation will attempt to gradually apply these displacements in $70$ equal increments, occasionally breaking some increments into finer steps in order to ensure convergence of the force-balancing algorithm. 
\begin{table}[!t]
    \centering
    \begin{tabular}{c|c|c|c}
         \# & $\alpha$ & $\beta$  & $\foldAngle_0$ \\
         1 & 1.4 & 1.0 & 0.5 \\
         2 & 1.6 & 1.0 & 0.5 \\
         3 & 1.7 & 1.0 & 0.5 \\
         4 & 2.01 & 1.0 & 0.5 \\
         5 & 2.1 & 1.0 & 0.5 \\
         6 & 1.4 & 1.0 & 0.7 \\
         7 & 1.5 & 1.0 & 0.7 \\
         8 & 2.01 & 1.0 & 0.7
    \end{tabular}
    \caption{Origami reference state geometries for the nonlinear data displayed in main text Fig.~\ref{fig:main3}}
    \label{tab:nonlinearRunsConfig}
\end{table}
%

To generate the data in main text Fig.~\ref{fig:main4}, we replicate the approach to the data in Fig.~\ref{fig:main3} for a modified set of sheet configurations. Again the modulus of folding $\modFold$ is zero while the modulus of panel bending sets the energy scale $\modBend = 1$ and the modulus of panel stretching remains $\modStretch=10^6$ at the previous value. Here, numerical data is collected for systems with unit cell geometry defined by the angles $\foldAngle_0=0.5$, $\beta = 1.0$, and a series of different $\alpha = \{  1.4, 1.45, 1.5, 1.6, 1.7, 1.8, 2.01, 2.1, 2.2 \}$, generating the variety of lines, colored according to the reference state Poisson's ratio this implies. For each of these sheet configurations, data is run at the system sizes $N = \{10, 20, 30, 40 \}$ and each of these is run at each of the loading conditions below
\begin{table}[!t]
    \begin{subtable}[t]{1.0\linewidth}
    \centering
    \begin{tabular}{c|cccc}
        \quad & center-bottom & center-left & center-top & center-right  \\ \hline
        load 1 $\frac{1}{N}\vvec{u}$ & $\{0.00015, -0.001, 0.00075\}$ &  $\{0.0002, 0.0006, -0.0005\}$ &  $\{0.00035, -0.0002, 0.0006\}$ & $\{-0.0007, 0.0006, -0.00085\}$ \\
        load 2 $\frac{1}{N}\vvec{u}$ & $\{0.0002, 0.0006, -0.0005\}$ &  $\{0.00015, -0.001, 0.00075\}$ &  $\{0.00035, -0.0002, 0.0006\}$ & $\{-0.0007, 0.0006, -0.00085\}$ \\
        load 3 $\frac{1}{N}\vvec{u}$ & $\{0.0002, 0.0006, -0.0005\}$ &  $\{0.00035, -0.0002, 0.0006\}$ &  $\{0.00015, -0.001, 0.00075\}$ & $\{-0.0007, 0.0006, -0.00085\}$ \\
        load 4 $\frac{1}{N}\vvec{u}$ & $\{0.0002, 0.0006, -0.0005\}$ &  $\{0.00035, -0.0002, 0.0006\}$ &  $\{-0.0007, 0.0006, -0.00085\}$ & $\{0.00015, -0.001, 0.00075\}$         
    \end{tabular}
    \caption{Loads applied to vertices located at the center of each system edge}
    \end{subtable}
    \begin{subtable}[t]{1.0\linewidth}
    \centering
    \begin{tabular}{c|cccc}
        \quad & bottom-left & bottom-right & top-left & top-right  \\ \hline
        load 5 $\frac{1}{N}\vvec{u}$ & $\{0.00015, -0.001, 0.00075\}$ &  $\{0.0002, 0.0006, -0.0005\}$ &  $\{0.00035, -0.0002, 0.0006\}$ & $\{-0.0007, 0.0006, -0.00085\}$ \\
        load 6 $\frac{1}{N}\vvec{u}$ & $\{0.0002, 0.0006, -0.0005\}$ &  $\{0.00015, -0.001, 0.00075\}$ &  $\{0.00035, -0.0002, 0.0006\}$ & $\{-0.0007, 0.0006, -0.00085\}$ \\
        load 7 $\frac{1}{N}\vvec{u}$ & $\{0.0002, 0.0006, -0.0005\}$ &  $\{0.00035, -0.0002, 0.0006\}$ &  $\{0.00015, -0.001, 0.00075\}$ & $\{-0.0007, 0.0006, -0.00085\}$ \\
        load 8 $\frac{1}{N}\vvec{u}$ & $\{0.0002, 0.0006, -0.0005\}$ &  $\{0.00035, -0.0002, 0.0006\}$ &  $\{-0.0007, 0.0006, -0.00085\}$ & $\{0.00015, -0.001, 0.00075\}$         
    \end{tabular}
    \caption{Loads applied to vertices located at the system corners}
    \end{subtable}
    \caption{Displacement loading conditions used to obtain numerical data displayed in main text Figs.~\ref{fig:main4}\&\ref{fig:main5}}
    \label{tab:linearLoads}
\end{table}
which are sufficiently small loadings that this data corresponds with the mechanics of linear deformations. Note that the values listed represent the imposed displacement divided by the number of unit cells across $(N)$, thereby indicating the manner in which the loads are scaled with system size. However, while this system size feature in our code is important for the convergence of nonlinear loading, this has negligible effect for the linear data of main text Figs.~\ref{fig:main4}\&\ref{fig:main5}. Data in these figures is generated using the analysis methods in Sec.~\ref{app:analysis}, and averaging these results over the different loads in Table~\ref{tab:linearLoads}.

To generate the configuration data analyzed in main text Fig.~\ref{fig:main5} follows a very similar procedure as that for Fig.~\ref{fig:main4}, yet for a different series of parameters. In this case, only the system size $N=20$ unit cells across is considered and we probe instead the dependence on the panel stretching stiffness. The data points in main text Fig.~\ref{fig:main5} are then generated at each of the values $\modStretch = \{10^1, 10^2, 10^3, 10^4, 10^5, 10^6, 10^7, 10^8 \}$ and at each of the loads and  folding geometries from the above description of main text Fig.~\ref{fig:main4}. Again, averaging the fitting error data over the different loads generates the displayed values. The method of quantifying the error in the fitting is explained in Section.~\ref{app:softModeFitting} below.




%


\section{Analysis of Origami Sheets}\label{app:analysis}

Here we detail the methods to analyze the simulation data for a force-balanced origami sheet in both the nonlinear and linear regimes. This will include the hyperribbon error and the fitting to soft mode error. 

\subsection{Adherence to shape hyperribbon}\label{app:hyperribbon}

The main text includes quantification of the observed data obtained in Section.~\ref{app:simulations} via the quality of adherence of the sheet midplane shape to the predicted hyperribbon described in Section.~\ref{app:spatialSoftModes}. 
To quantify such adherence, we must first evaluate the fundamental forms of the sheet midplane which quantify the coarse shape locally. To do so, we follow previously employed methods for the evaluation of nonlinear deformation in lattice metamaterials~\cite{Czajkowski2022, Czajkowski2022-2}. We begin with the evaluation of the metric.  As the lattice system is spatially discrete, we do not directly measure a continuous tensor field, but rather estimate local tensor values at each unit cell grid index $(n_1, n_2)$. As described in the beginning of Sec.~\ref{app:coarseModes}, the metric of deformation for an initially uniform lattice may be estimated using the reference (initial) $\init{\latticeVec{1}}, \init{\latticeVec{2}}$ and deformed (final) $\latticeVec{1}, \latticeVec{2}$ lattice vectors via Eq.~\ref{eq:app_metricGeneral}. To extract the lattice vectors $\latticeVec{1}(n_1, n_2), \latticeVec{2}(n_1, n_2)$ at the arbitrary unit cell index $(n_1, n_2)$, a common and simple choice is to use the vector which connects the position of the $\vertIndex$-th node $\posr_{\vertIndex}(n_1, n_2)$ in this unit cell to the position of its counterpart in the next unit cell $\posr_{\vertIndex}(n_1 + 1, n_2)$.  
However, the asymmetry in such a choice yields error in the estimation of the metric at the order of the lattice lengths themselves and the gradients of the metric. Conveniently, a symmetric choice to define these lattice vectors, averaging ``backward'' and ``forward'' estimates of such together, eliminates this error in favor of smaller, higher order errors. In addition, as noted in Ref.~\cite{mcinerney2022discrete}, these lattice vectors must be averaged over the nodes of the unit cell to avoid spurious symmetry-violating extra terms in the estimation of these fundamental forms. With this, our formulas for extraction of the lattice vectors from the numerical data are
\begin{align}\label{eq:app_latticeVecExtractOne}
    \latticeVec{1}(n_1, n_2) & \sim \frac{1}{4} \sum_{\vertIndex \in \text{unit cell}} \frac{1}{2} \left( \posr_{\vertIndex}(n_1+1, n_2) - \posr_{\vertIndex}(n_1-1, n_2)  \right) \\ \label{eq:app_latticeVecExtractTwo}
    \latticeVec{2}(n_1, n_2) & \sim \frac{1}{4} \sum_{\vertIndex \in \text{unit cell}} \frac{1}{2} \left( \posr_{\vertIndex}(n_1, n_2+1) - \posr_{\vertIndex}(n_1, n_2-1)  \right) \, .
\end{align}
With this, relying on the definition established in Eq.~\ref{eq:app_metricGeneral}, the metric
\begin{equation}\label{eq:app_metric_lattice}
\data{\metric}_{ij}(n_1, n_2) = \frac{\latticeVec{i}(n_1, n_2) \cdot \latticeVec{j}(n_1, n_2)}{|\init{\latticeVec{i}}| |\init{\latticeVec{j}}|}
\end{equation}
is readily extracted for each unit cell in the simulation data; and similarly for the strain using Eq.~\ref{eq:app_strainDef}. Here, the superscript indicates that this is the observed value of the metric, so estimated from the data.  

The lattice vectors extracted using the recipes in Eqs.~\ref{eq:app_latticeVecExtractOne}\&\ref{eq:app_latticeVecExtractTwo} are also used to estimate the local value of the curvature tensor, and again the symmetric definition averaged over the unit cell eliminates significant leading error. First the lattice vectors are used to first define the local normal of the smoothed sheet
\begin{equation}\label{eq:normalFromLattiveVecs}
    \normal(n_1, n_2) = \frac{\latticeVec{1}(n_1, n_2)\times \latticeVec{2}(n_1, n_2)}{|\latticeVec{1}(n_1, n_2)\times \latticeVec{2}(n_1, n_2)|}
\end{equation}
which is useful in defining the curvature. Further, the gradient of this normal with respect to the sheet parametrization coordinates must be known. This is achieved by again using symmetric derivatives so that we write
\begin{align}\label{eq:app_normalGradient}
    \partial_1 \normal(n_1, n_2) & = \frac{1}{2}(\normal(n_1+1, n_2) - \normal(n_1-1, n_2))
    \partial_2 \normal(n_1, n_2) & = \frac{1}{2}(\normal(n_1, n_2+1) - \normal(n_1, n_2-1)) \, .
\end{align}
With this, the recipe for the curvature tensor components 
\begin{equation}\label{eq:app_extractCurvature}
\curvature_{ij} = \begin{bmatrix} -\latticeVec{1} \cdot \partial_1 \normal & -\latticeVec{2}\cdot\partial_1\normal \\ -\latticeVec{1}\cdot\partial_2\normal & -\latticeVec{2}\cdot\partial_2\normal  \end{bmatrix}
\end{equation}
where the dependence on the unit cell index location is now suppressed for convenience, and index (lhs) versus matrix (rhs) representation are mixed for simplicity of notation. 

Once the fundamental forms are determined across the origami sheet, we must determine the quality of adherence of these tensors to the hyperribbon. This is achieved by first obtaining the closest value of the folding parameter $\foldAngle$ and then evaluating the twist and bend components of the curvature tensor given this value of $\foldAngle$. The folding angle is determined most readily using the local value of $\latticeVec{2}$. In the case of a uniformly applied folding motion, this lattice vector length is determined by the formula $|\latticeVec{2}| = 2 a \cos(\foldAngle)$ as in Section.~\ref{app:foldPatterns} Eq.~\ref{eq:app_latticeVecMorphTwo}. Inverting this equation for the folding angle gives a sufficiently accurate estimate for the folding angle. 

Having obtained the folding angle, it is possible to then also determine the twist and bend mode amplitudes locally. The local twist mode is readily obtained as the only off-diagonal component of the curvature tensor, so that $\twist(n_1, n_2) = \curvature_{12}(n_1, n_2)$. Note that using the other off-diagonal component is equally valid $\twist(n_1, n_2) = \curvature_{21}(n_1, n_2)$ and in practice, the average pf these two estimates is employed to remove bias. Extracting the bend component is similarly achievable using either of the diagonal components of curvature $\curvature_{11}$ or $\curvature_{22}$. Explicitly, either of the formulas
\begin{align}\nonumber
    \bend(n_1, n_2) & = \frac{\curvature_{11}(n_1, n_2)}{\principalOne(\foldAngle(n_1, n_2)) \principalOne'(\foldAngle(n_1, n_2))} \\ \nonumber
    \bend(n_1, n_2) & = -\frac{\curvature_{22}(n_1, n_2)}{\principalTwo(\foldAngle(n_1, n_2)) \principalTwo'(\foldAngle(n_1, n_2))}
\end{align}
where $\principalOne$ and $\principalTwo$ are the known principal stretch functions introduced in Sec.~\ref{app:linearModesDefine}. Again, in practice each choice of estimation method is roughly equivalent and the two estimates are averaged to remove bias.

Finally, having obtained local estimates of the soft mode fields, we can quantify the accuracy of this hyperribbon description. To do this, we then produce local estimates of the projection of the fundamental forms onto the hyperribbon $\projection{\metric}_{ij}(n_1, n_2)$ and $ \projection{\curvature}_{ij}(n_1, n_2)$ at each unit cell.  This is achieved by reconstructing the fundamental forms from the soft mode field data using the discrete version of Eqs.~\ref{eq:app_metricFieldAssumption}\&\ref{eq:app_curvatureFieldAssumption}
\begin{align}\label{eq:app_reconstructMetric}
    \projection{\metric}_{ij}(n_1, n_2) & = \metric^*_{ij}(\foldAngle(n_1, n_2)) \\ \label{eq:app_reconstructCurvature}
    \projection{\curvature}_{ij}(n_1, n_2) & = \curvature^*_{ij}(\foldAngle(n_1, n_2), \twist(n_1, n_2), \bend(n_1, n_2)) \, ,
\end{align}
where the functions $\metric^*_{ij}$ and $\curvature^*_{ij}$ are defined in Sec.~\ref{app:linearModesDefine} to capture the hyperribbon. With this, the normalized tensor error in either the metric or the curvature may (first) be readily constructed from these two tensor fields via the tensor error function
\begin{equation}\label{eq:tensorFieldError}
    \Delta[t^{(1)}_{ij}, t^{(2)}_{ij}] = \sqrt{\frac{\sum_{n_1, n_2} \sum_{ij} (t^{(1)}_{ij}(n_1, n_2) - t^{(2)}_{ij}(n_1, n_2))^2}{\sum_{n_1, n_2} \sum_{ij}  t^{(1)}_{ij}(n_1, n_2)t^{(1)}_{ij}(n_1, n_2)}} \, .
\end{equation}
which appropriately yields the fractional difference between two discrete tensor grids $t^{(2)}_{ij}(n_1, n_2)$ and $t^{(2)}_{ij}(n_1, n_2)$ which are close in value. 
It is important to notice that this error evaluation method is most appropriate for tensors which are zero in the undeformed state. Otherwise, the quality of error is artificially inflated for small deformations. The curvature tensor satisfies this condition, and therefore the appropriate dimensionless measure of the curvature portion of the hyperribbon error will be $\Delta_{\curvature} = \Delta[\data{\curvature}_{ij}, \projection{\curvature}_{ij}]$. However, to quantify the fractional non-hyperribbon component of the metric we instead use the deviational part of this tensor, which is the strain. The strain is constructed by $\strain_{ij} = \frac{1}{2} \left( \metric_{ij} - \metric_{ij}^{(0)} \right)$, where the initial metric $\metric_{ij}^{(0)}$ in the case here is the identity $\delta_{ij}$. Then this portion of the hyperribbon error corresponding to the metric is $\Delta_{\strain} = \Delta\left[\frac{1}{2}\left(\data{\metric}_{ij} - \delta_{ij} \right), \frac{1}{2}\left(\projection{\metric}_{ij} - \delta_{ij} \right)\right]$.


\subsection{Analytic soft mode fitting}
\label{app:softModeFitting}

The theory presented in the main text includes not only the assertion the local fundamental forms be characterized by the three mode amplitudes (hyperribbon), but also prescribes the manner in which these modes will be distributed across the sheet spatially. In the limit of linear deformations, we indicate that these modes must be controlled by sheared analytic functions, detailed in Sec.~\ref{app:linearAnalytics}. In order to evaluate this hypothesis, in this section we present an analysis procedure used to take a sheet geometric configuration and extract a single dimensionless number describing how close the state is to such a sheared analytic mode. This method is used to produce the data displayed in main text Fig.~\ref{fig:main5}. As with the above hyperribbon error analysis, this is performed separately for the strain and curvature portions of the fundamental form variations. Note that this analysis closely follows that established in Ref.~\cite{Czajkowski2022-2}.

This fitting procedure first follows the analysis in Sec.~\ref{app:hyperribbon} to obtain a numeric estimate of the local mode values $\foldAngle(n_1, n_2), \twist(n_1, n_2), \bend(n_1, n_2)$ for each unit cell grid index $(n_1, n_2)$ in the sheet. Each unit cell is affiliated with a location
\begin{equation}\label{eq:app_unitCellLocation}
    \posr(n_1, n_2) \equiv \frac{1}{4} \sum_{\vertIndex \in \begin{smallmatrix}\text{unit} \\ \text{cell} \end{smallmatrix}} \posr_{\vertIndex}(n_1, n_2)
\end{equation}
by summing over the four vertex locations within the unit cell, again similar to that described in the previous section for the lattice vectors. Here, $ \posr_{\vertIndex}(n_1, n_2)$ is the location of the $\vertIndex$-th of the 4 vertices of the $(n_1, n_2)$ unit cell. Equipped with a sufficient set of field and position data, we set out to fit each field individually to an appropriate analytic form. As detailed in Section~\ref{app:linearAnalytics}, linear combinations of the curvature fields,
\begin{align}\label{eq:app_curvatureShearedAnalytic}
    \curvatureVar & = \twist - \gamgam  \principalTwo' \bend = \curvatureModeOne(\ww) \\ \label{eq:app_curvatureBarShearedAnalytic}
    \bar{\curvatureVar} & = \twist + \gamgam  \principalTwo' \bend = \curvatureModeTwo(\bar{\ww}) \, ,
\end{align}
will be captured by analytic functions of the sheared reference-frame spatial variables $\ww, \bar{\ww}$ (termed ``sheared analytic functions'') for the lowest-energy deformations. Linear combinations of the strain variables are captured similarly
\begin{align}\label{eq:app_strainShearedAnalytic}
    \strainVar & = \aA \mechStrain - \gamgam \coarseRot = \strainModeOne(\ww) \\ \label{eq:app_strainBarShearedAnalytic}
    \bar{\strainVar} & = \aA \mechStrain + \gamgam \coarseRot = \strainModeTwo(\bar{\ww}) \, .
\end{align}
Values for these spatial variables $\{\ww(n_1, n_2), \bar{\ww}(n_1, n_2)\}$ are easily obtained from the initial unit cell position data $\{\init{\posr}(n_1, n_2) \}$ via the recipes $\ww(\init{\posr}) = \init{\posr} \cdot \init{\latticeVec{1}} + \sqrt{\poissonIn}\, \init{\posr} \cdot \init{\latticeVec{2}}$ and $\bar{\ww}(\init{\posr}) = \init{\posr} \cdot \init{\latticeVec{1}} - \sqrt{\poissonIn}\, \init{\posr} \cdot \init{\latticeVec{2}}$.
Here, we borrow from the local basis notation constructed using the lattice vector directions in the previous section (\ref{app:hyperribbon}). Choosing the coordinate system so that the origami sheet midplane initially lies in the x-y plane, these may be pictured as the x and y spatial components.  Note that because these are reference frame coordinates they do not change with the sheet deformation, and are always defined using  $\{\init{\posr}(n_1, n_2) \}$.

Solving directly for the mechanism strain amplitude $\mechStrain$ in Eqs.~\ref{eq:app_strainShearedAnalytic}\&\ref{eq:app_strainBarShearedAnalytic} we see that this will be captured by a particular analytic form 
\begin{equation}\label{eq:app_mechStrainAnalytic}
\target{\mechStrain}(\ww, \bar{\ww}) = \frac{1}{2 \aA}(\strainModeOne(\ww) + \strainModeTwo(\bar{\ww})).
\end{equation}
Ignoring the rotation field $\coarseRot$, this captures the essential restriction on the possible spatial forms that this mechanism strain may take through space. The metric of deformation is independent of this rotation field and evaluating the adherence of the first fundamental form variations to  our predictions is accomplished by fitting the spatial data for this mechanism strain to the form above. This procedure appears different on each side of the exceptional point at $\poissonIn=0$ and it is therefore useful to include these details more carefully. We consider the case of the mechanism strain pattern first, with the procedure for the twist and bend fields following a nearly identical procedure as discussed near the end of this section. In what follows we will absorb the factor of $\frac{1}{2 \aA}$ into the definition of the strain modes, so that the procedure applies equally to each of the modes we will fit $(\mechStrain, \twist, \bend)$ which are each fit to the form $(\strainModeOne(\ww)+\strainModeTwo(\bar{\ww}))$. 

\subsubsection{Auxetic Case}

For origami sheets with auxetic mechanism $\poissonIn<0$, the quantity $\gamgam = \sqrt{1/\poissonIn}$ is complex-valued, and so are the position data $\{\ww(n_1, n_2), \bar{\ww(n_1, n_2)}\}$. Therefore, analytic functions $\strainModeOne, \strainModeTwo$ will necessarily also be complex valued. The combination of these two functions to produce the local mechanism strain amplitude, however, is required to be real-valued. Therefore, in the auxetic case it will be required that these functions are complex conjugates of one another. As these functions admit expansions,
\begin{align}\label{eq:app_expandAuxAnalyticStrainOne}
    \strainModeOne(\ww) & = \sum_{n=0}^{N_c} C_n \ww^n \\ \label{eq:app_expandAuxAnalyticStrainTwo}
    \strainModeTwo(\bar{\ww}) & = \sum_{n=0}^{N_c} D_n \bar{\ww}^n
\end{align}
and since the bar operation on $\ww$ is here equivalent to complex conjugation, the requirement that $\mechStrain$ be real-valued is captured by $C_n = \cc{D}_n$. Therefore, finding the closest fit for the mechanism strain data is a process of varying the $C_n$ complex coefficients to minimize the fitting error function
\begin{equation}
    \delta_{\text{fit}}[\{\mechStrain^{(k)} \}, \target{\mechStrain}()] = \sum_k \left|\mechStrain^{(k)} - \target{\mechStrain}(\ww_k, \bar{\ww}_k)  \right|^2
\end{equation}
where all interior unit cells are now indexed by a single number $k$ so that $\posr_k = \posr(n_1(k), n_2(k))$, and again a boundary layer of unit cells which are two or less away from the system edge are excluded from the analysis to minimize such boundary effects.
For the data explored here, we choose to have $N_c = 20$. 

To solve for the coefficients which minimize the fitting error function, we first break down into real and imaginary parts $C_n = A_n + i B_n$. In this case the function becomes
\begin{equation}
  \delta_{\text{fit}} = \sum_k \left[ \mechStrain^{(k)} - \sum_n (2 A_n \mathrm{Re}[\ww_k^n] - 2 B_n \mathrm{Im}[\ww_k^n]) \right]^2
\end{equation}
and the minimizing coefficients are then the solution of the set of equations
\begin{align}
  \frac{\partial \delta_{\text{fit}}}{\partial A_m} = & 0 = -2 \sum_k\left[ \mechStrain^{(k)} - \sum_n (2 A_n \mathrm{Re}[\ww_k^n] - 2 B_n \mathrm{Im}[\ww_k^n]) \right] 2 \mathrm{Re}[\ww_k^m] \\
  \frac{\partial \delta_{\text{fit}}}{\partial B_m} = & 0 = 2\sum_k\left[ \mechStrain^{(k)} - \sum_n (2 A_n \mathrm{Re}[\ww_k^n] - 2 B_n \mathrm{Im}[\ww_k^n]) \right] 2 \mathrm{Im}[\ww_k^m]
\end{align}
which must be satisfied simultaneously for every $m$. This is a linear algebra problem which may be put conveniently into a matrix form
\begin{equation}\label{eq:app_aux_fitting_linearAlgebra}
  \begin{bmatrix}
    \vvec{f}_1 \\ \vvec{f}_2
  \end{bmatrix}
   =
    \begin{bmatrix}
      \ttens{M}_{1A} & \ttens{M}_{1B} \\
      \ttens{M}_{2A} & \ttens{M}_{2B}
    \end{bmatrix}
     \cdot
  \begin{bmatrix}
    \vvec{A} \\
    \vvec{B}
  \end{bmatrix}.
\end{equation}
Here, $\vvec{f}_1$ and $\vvec{f}_2$ are vectors having components $(f_1)_m = \sum_k\mechStrain^{(k)} \mathrm{Re}[\ww_k^m] $ for $m={0,1,2,...N_c-1}$ and $(f_2)_m = -\sum_k\mechStrain^{(k)} \mathrm{Im}[\ww_k^m] $ for $m={1,2,...N_c-1}$, while $\ttens{M}_{1A}, \ttens{M}_{2A}, \ttens{M}_{1B}, \ttens{M}_{2B}$ are matrices defined by
\begin{align}
(\ttens{M}_{1A})_{mn} & =  2\sum_k \mathrm{Re}
[\ww_{(k)}^m] \mathrm{Re}[\ww_{(k)}^n] \\
& m=\{ 0, 1, ... N_c-1 \} \quad n=\{ 0, 1, ... N_c-1 \} \\
(\ttens{M}_{2A})_{mn} & =  2\sum_k \mathrm{Im}[\ww_{(k)}^m] \mathrm{Re}[\ww_{(k)}^n] \\
& m=\{ 1, 2, ... N_c-1 \} \quad n=\{ 0, 1, ... N_c-1 \} \\
(\ttens{M}_{1B})_{mn} & =  2\sum_k \mathrm{Re}[\ww_{(k)}^m] \mathrm{Im}[\ww_{(k)}^n] \\
& m=\{ 0, 1, ... N_c-1 \} \quad n=\{ 1, 2,... N_c-1 \} \\
(\ttens{M}_{2B})_{mn} & =  2\sum_k \mathrm{Im}[\ww_{(k)}^m] \mathrm{Im}[\ww_{(k)}^n] \\
& m=\{ 1, 2, ... N_c-1 \} \quad n=\{ 1, 2, ... N_c-1 \} 
\end{align}
and the vectors of coefficients are
\begin{align}
\vvec{A}^T & = \{ A_0, A_1, ... A_{N_c-1} \} \\
\vvec{B}^T & = \{ B_1, B_2, ... B_{N_c-1} \} .
\end{align}
Note that the element $B_0$ has no impact on the values of the function $\target{\mechStrain}(\ww, \bar{\ww})$, and we have already omitted the corresponding row and column from the elements in Eq.~\ref{eq:app_aux_fitting_linearAlgebra}  leaving $B_0$ undetermined, and avoiding numerical issues with solving the linear algebra problem. This linear system of equations (Eq.~\ref{eq:app_aux_fitting_linearAlgebra}) is solved in Mathematica, yielding the coefficients $C_n$ that define the closest fit function. 

The above procedure obtains the closest fit sheared analytic mode matching to the mechanism strain field $\target{\mechStrain}()$. The procedure is carried out identically to procure predictive fittings to the twist and the bend fields each, respectively yielding the closest fit functions $\target{\twist}()$ and $\target{\bend}()$. It is worth noting that this procedure does not as carefully probe that the combination field $\curvatureVar \equiv \twist - \sqrt{\frac{1}{\poissonIn}} \principalTwo'\bend$ precisely adheres to a single sheared analytic mode. Rather, this process confirms that the twist and bend each individually obey the conditions entailed by the sheared analytic solutions. This is like checking a complex function for analyticity by instead checking that the real and imaginary components each obeyed the laplace equation, without checking the harmonic conjugacy directly. Further investigations may improve upon this method. However, the quality of the data and other informal measures indicate that these fields indeed obey the sheared analytic generalization of harmonic conjugacy. 

Given these closest fit functions $\target{\mechStrain}(), \target{\bend}(), \target{\twist}()$, we would like to now evaluate how close the estimates are to the target data they are intended to approximate. This is accomplished similar to in the previous section, by first extracting the corresponding strain 
\begin{equation}
\strain^{(\text{pred})}_{ij}(n_1, n_2) = \target{\mechStrain}\left(\ww(n_1, n_2), \bar{\ww}(n_1, n_2)\right) \aA \begin{bmatrix} 1 & 0 \\ 0 & -\poissonIn \end{bmatrix}
\end{equation}
and the corresponding curvature 
\begin{equation}
\curvature^{(\text{pred})}_{ij}(n_1, n_2) = \target{\twist}\left(\ww(n_1, n_2), \bar{\ww}(n_1, n_2)\right) \begin{bmatrix} 0 & 1 \\ 1 & 0 \end{bmatrix} + \target{\bend}\left(\ww(n_1, n_2), \bar{\ww}(n_1, n_2)\right) \begin{bmatrix} \bend_1(\init{\foldAngle}) & 0 \\ 0 & \bend_2(\init{\foldAngle}) \end{bmatrix}.
\end{equation}
Then the error in the fitting for the strain is accomplished by reusing the error function (Eq.~\ref{eq:tensorFieldError}) from Section~\ref{app:hyperribbon}, this time using the analytic predictions of the tensors rather than the projections into the hyperribbon. For the case of the curvature, then, the fractional error in the analytic fitting is $\Delta^{(fit)}_{\curvature} = \Delta[\curvature^{(\text{pred})}_{ij}, \data{\curvature}_{ij}]$, which takes two grids of tensors and reduces them to their root-mean-square difference normalized by their root-mean-square projection onto one another. Similarly, this fractional error in fitting the strain is $\Delta^{(fit)}_{\strain} = \Delta[\strain^{(\text{pred})}_{ij}, \data{\strain}_{ij}]$, where $\data{\strain}_{ij} = \frac{1}{2}\left( \data{\metric}_{ij} - \delta_{ij} \right)$.

\subsubsection{Anauxetic Case}
The procedure for closest fitting to the soft modes in the case of the anauxetic mechanism is distinct yet quite similar. The important distinction comes from 
$\gamgam = \sqrt{\frac{1}{\poissonIn}}$ which is now real-valued; therefore the coordinates $\ww$ and $\bar{\ww}$ are also real-valued everywhere. Therefore instead of the inter-relation between complex coefficients of the analytic expansions
\begin{align}\label{eq:app_expandAnauxAnalyticStrainOne}
    \strainModeOne(\ww) & = \sum_{n=0}^{N_c} C_n \ww^n \\ \label{eq:app_expandAnauxAnalyticStrainTwo}
    \strainModeTwo(\bar{\ww}) & = \sum_{n=0}^{N_c} D_n \bar{\ww}^n
\end{align}
we instead have real and independent coefficients $\{C_n\}$ and $\{D_n\}$. Similar to the auxetic case, we again minimize the fitting error function, which now takes the form
\begin{equation}
  \delta_{\text{fit}} = \sum_k \left[ \mechStrain^{(k)} - \sum_n  C_n \ww_k^n  - \sum_n  D_n \bar{\ww}_k^n  \right]^2 .
\end{equation}
We then take partial derivatives to minimize, again leading to a set of linear equations, captured in matrix form
\begin{equation}\label{eq:app_anaux_fitting_linearAlgebra}
  \begin{bmatrix}
    \vvec{f}_1 \\ \vvec{f}_2
  \end{bmatrix}
   =
    \begin{bmatrix}
      \ttens{M}_{1C} & \ttens{M}_{1D} \\
      \ttens{M}_{2C} & \ttens{M}_{2D}
    \end{bmatrix}
     \cdot
  \begin{bmatrix}
    \vvec{C} \\
    \vvec{D}
  \end{bmatrix}.
\end{equation}
Here, $\vvec{f}_1$ and $\vvec{f}_2$ are vectors having components $(f_1)_m = \sum_k\mechStrain^{(k)} \mathrm{Re}[\ww_k^m] $ for $m={0,1,2,...N_c-1}$ and $(f_2)_m = -\sum_k\mechStrain^{(k)} \mathrm{Im}[\ww_k^m] $ for $m={1,2,...N_c-1}$, while $\ttens{M}_{1A}, \ttens{M}_{2A}, \ttens{M}_{1B}, \ttens{M}_{2B}$ are matrices defined by
\begin{align}
(\ttens{M}_{1C})_{mn} & =  \sum_k \ww_{(k)}^m \ww_{(k)}^n \\
& m=\{ 0, 1, ... N_c-1 \} \quad n=\{ 0, 1, ... N_c-1 \} \\
(\ttens{M}_{2C})_{mn} & =  \sum_k \bar{\ww}_{(k)}^m \ww_{(k)}^n \\
& m=\{ 1, 2, ... N_c-1 \} \quad n=\{ 0, 1, ... N_c-1 \} \\
(\ttens{M}_{1D})_{mn} & =  \sum_k \ww_{(k)}^m \bar{\ww}_{(k)}^n \\
& m=\{ 0, 1, ... N_c-1 \} \quad n=\{ 1, 2,... N_c-1 \} \\
(\ttens{M}_{2D})_{mn} & =  \sum_k \bar{\ww}_{(k)}^m \bar{\ww}_{(k)}^n \\
& m=\{ 1, 2, ... N_c-1 \} \quad n=\{ 1, 2, ... N_c-1 \} 
\end{align}
and the vectors of coefficients are
\begin{align}
\vvec{C}^T & = \{ C_0, C_1, ... C_{N_c-1} \} \\
\vvec{D}^T & = \{ D_1, D_2, ... D_{N_c-1} \} .
\end{align}
Again, one element and row have been omitted to account for the single indeterminacy. In this case it is because $C_0$ and $D_0$ are redundant in determining the uniform part of the resulting soft mode. In this case it is a choice to omit $D_0$. 

Solving these equations numerically in Mathematica yields the prediction for the modes, and analyzing the fractional error in this fitting follows the same procedure as the auxetic case, leading to the data in main text Fig.~\ref{fig:main5}. 

\bibliography{bibliofile}
\end{document}